\documentclass[aps,prl,reprint,superscriptaddress]{revtex4-1}
\usepackage[dvipsnames]{xcolor}
\usepackage{graphicx}
\usepackage{amsmath,amssymb}
\usepackage{bm}
\usepackage{braket}
\usepackage{lipsum}
\usepackage{makerobust}
\usepackage{simpler-wick}
\usepackage{hyperref}

\newcommand*\circled[1]{\tikz[baseline=(char.base)]{
    \node[shape=circle,draw,inner sep=1pt] (char) {#1};}}
\MakeRobustCommand\circled

\newcommand{\makeauthor}[2]{\newcommand{#1}[1]{{%
  \sffamily\color{#2}{%
    \bfseries\begingroup\escapechar=-1\edef\x{\endgroup\string#1}\x:%
  } ##1}}%
  \MakeRobustCommand#1}
\makeauthor{\eric}{Plum}
\makeauthor{\themba}{ForestGreen} 
\makeauthor{\dc}{magenta}
\makeauthor{\sr}{blue}
\makeauthor{\Fig}{red}

\begin{document}

\renewcommand{\vec}[1]{\bm{#1}}
\newcommand{\up}{{\uparrow}}
\newcommand{\dw}{{\downarrow}}
\newcommand{\pa}{{\partial}}
\newcommand{\pd}{{\phantom{\dagger}}}
\newcommand{\bs}[1]{\boldsymbol{#1}}
\newcommand{\add}[1]{{{\color{blue}#1}}}
\newcommand{\addn}[1]{{{\color{red}#1}}}
\newcommand{\todo}[1]{{\textbf{\color{red}ToDo: #1}}}
\newcommand{\tbr}[1]{{\textbf{\color{red}\underline{ToBeRemoved:} #1}}}
\newcommand{\eps}{{\varepsilon}}
\newcommand{\nn}{\nonumber}

\newcommand{\brap}[1]{{\bra{#1}_{\rm phys}}}
\newcommand{\bral}[1]{{\bra{#1}_{\rm log}}}
\newcommand{\ketp}[1]{{\ket{#1}_{\rm phys}}}
\newcommand{\ketl}[1]{{\ket{#1}_{\rm log}}}
\newcommand{\braketp}[1]{{\braket{#1}_{\rm phys}}}
\newcommand{\braketl}[1]{{\braket{#1}_{\rm log}}}

\graphicspath{{./}{./figures/}}



\title{Characterizing Dynamic Majorana Hybridization\\
for Universal Quantum Computing}

\author{Themba Hodge}
\author{Eric Mascot}
\author{Dan Crawford}
\affiliation{School of Physics, University of Melbourne, Parkville, VIC 3010, Australia}
\author{Stephan Rachel}
\affiliation{School of Physics, University of Melbourne, Parkville, VIC 3010, Australia}
\noaffiliation

\date{\today}

\begin{abstract}
Qubits built out of Majorana zero modes (MZMs) have long been theorized as a potential pathway toward fault-tolerant topological quantum computation. 
Almost unavoidable in these processes is 
Majorana wavefunction overlap, known as hybridization, which arise throughout the process when Majorana modes get close to each other. This breaks the ground state degeneracy, leading to qubit errors in the braiding process.
This work presents a simple but precise method to predict qubit errors for dynamic hybridization which varies in space and time. This includes hybridization between four or more MZMs through topological or  trivial regions, or both of them.
%
As an illustration, we characterize qubit-errors for an X-gate.
We demonstrate how to utilize the hybridization to implement not only arbitrary one-qubit rotations but also a two-qubit {\it controlled} variable phase gate,  providing a demonstration of universal quantum computing.
\end{abstract}

\maketitle

%
%

{\it Introduction.---} Topological quantum computing presents one of the most exciting 
avenues towards fault-tolerant quantum computation\,\cite{kitaev2003,nayak2008}. 
In essence, this depends on the storing and manipulation of quantum information non-locally in the form of \emph{Majorana zero modes} (MZMs), zero energy subgap states that form on the boundaries of topological superconductors\,\cite{kitaev2001,read-00prb10267}. 
Promising platforms to host these exotic states 
have been explored\,\cite{lutchyn-10prl077001,oreg-10prl177002,Deng16,mourik2012,microsoft23,nadjperge13,Pientka2013,Braunecker2013,Vazifeh2013,Klinovaja2013,pawlak2016,palacio2019,schneider21}.
%
MZMs obey non-Abelian statistics under exchange, otherwise known as a \emph{braid}. Such an operation changes their quantum state, allowing for the encoding of unitary gates \cite{bravyi2002}. 
Unlike most quantum computing platforms, non-locality protects the topological quantum bits and gates 
from the effects of local perturbation, providing an exciting avenue towards 
\emph{fault-tolerant quantum computing}\,\cite{nayak2008,wang2010,dassarma-15npjqi15001,lahtinen-17spp021}.

While true MZMs remain at zero energy, in any finite, realistic system Majorana wavefunction overlap, known as \emph{hybridization}, will perturb these states to finite energies\,\cite{cheng2009}. 
With the advent of 
superconducting nano-structures protected by small bulk energy gaps\,\cite{Li2014,li2016,pahomi2020,crawford2022,Li2023}
avoiding the overlap of MZMs in experiment becomes almost unavoidable. Previous works have reported oscillations in the fidelity of a braid\,\cite{cheng2011,Sozinho2015,bauer2018,harper2019,Chen2022,bedow2024}, however, a precise dynamical understanding of why this occurs remains unknown.
Further, braiding MZMs does not lead to universal quantum computing\,\cite{bravyi2002}.
%
We require some controlled hybridization protocol to implement the $\pi/4$-phase gate or \emph{T-gate}\,\cite{Freedman2006,Bonderson2010}, the missing piece for universal quantum computation based on MZMs. Previously, several clever ideas to realize a phase gate were proposed such as 
using geometric phases\,\cite{Karzig2016}, measurements\,\cite{Karzig2019,Clarke2016}, magic states\,\cite{Litinski2018,OBrien2018}, and hybrid qubit designs\,\cite{bonderson2011,Jiang2011}.
Any proposal requires, however,
a complete understanding of how hybridization changes the many-body ground state.
This is pivotal: 
(i) To well define the dynamics of MZM hybridization over a braiding process, and thus the accumulation of error in the ground state manifold.
(ii) To accurately predict the timescales required for hybridization based quantum gates based.

In this Letter, we quantitatively predict the effects of hybridization in dynamical braiding processes.
Such methods have been utilized previously for a pair of vortices in a $p$-wave superconductor with fixed distance, corresponding to a pair of MZMs with fixed energy splitting\,\cite{cheng2011} and to investigate non-adiabatic transitions to bulk states in Majorana exchange processes\,\cite{scheurer2013}. Here we characterize the error for systems with multiple pairs of MZMs and time-dependent energy splitting.   
We also predict and simulate the T-gate or magic gate\,\cite{Bonderson2010,Karzig2016,Karzig2019,Litinski2018,OBrien2018,Clarke2016,bonderson2011,Jiang2011}, along with an arbitrary phase gate along the $y$-axis.
In contrast to previous work, our focus is on quantifying the dynamic hybridization in a many-MZM system and on
demonstrating how it 
may be utilized to map an arbitrary state about the Bloch sphere.
Finally, we extend the scope of our method to implement an arbitrary \emph{controlled-phase} gate, unattainable via usual braiding procedures.

%
%

{\it Method.---}
We consider a time-dependent 
Hamiltonian $\mathcal{H}(t)=\frac{1}{2}
    \hat{\Psi}^\dag
    H_{\textrm{BdG}}(t) 
    \hat{\Psi}$ with the Bogoliubov--de Gennes (BdG) matrix 
\begin{equation}
    H_{\textrm{BdG}}(t) =
    \begin{pmatrix}
        h(t) & \Delta(t)\\[10pt] 
        \Delta^{\dag}(t) & -h^T(t)
    \end{pmatrix}
\label{eq:MFHam}
\end{equation}

where $\hat{\Psi}=(\hat{c}_1,...,\hat{c}_N,\hat{c}^{\dag}_{1},...,\hat{c}^{\dag}_N)$. The $c_n$ ($c^{\dag}_n$) are the annihilation (creation) operators for a bare electron with index $n$ (e.g. position, spin etc.). 
We keep this general, as it may label multiple physical parameters such as the site, spin and orbital of the electron. 
$h_{mn}(t)$ is a time-dependent hopping between states $m$ and $n$, and $\Delta_{mn}(t)$ is the mean field pairing potential.
We may transform between superconducting quasiparticle operators $\{\hat{d}_\mu\}$ and electron operators $\{\hat{c}_a\}$ via a unitary \emph{Boguliubov transformation}, given by $\hat{d}^{\dag}_\mu=\hat{c}^{\dag}_aU_{a\mu}+\hat{c}_aV_{a\mu}$ for some $U_{a\mu}, V_{a\mu} \in \mathbb{C}$. 
This is such that $\mathcal{H}(0)=\sum_\mu E_\mu\hat{d}^{\dag}_\mu\hat{d}_\mu-\frac{1}{2}\sum_\mu E_\mu$, with $E_\mu \geq 0$ the eigenenergies of the system.
To track the time-evolution of the state-space, we consider the time-evolution of the superconducting quasiparticles $\hat{d}_\mu(t)=\mathcal{S}(t,t_0)\hat{d}_\mu(t_0)\mathcal{S}^{\dag}(t,t_0)$, where $\mathcal{S}(t,t_0)$ is the time-evolution operator. 
It is defined as $\mathcal{S}(t,t_0)=\mathcal{T}\textrm{exp}(-i\int^t_{t_0}\mathcal{H}(t')dt')$, with $\mathcal{T}$ time-ordering the matrix exponential. 
The $\hat{d}^{\dag}_\mu(t)$ create the dynamical single-particle state $|\Psi_\mu(t)\rangle$ by acting on the time-evolved quasiparticle vacuum, i.e., $\hat{d}^{\dag}_\mu(t)|0(t)\rangle_d=|\Psi_\mu(t)\rangle$, with Boguliubov wavefunction $\Psi_\mu(t)$. 
We may represent this as an expansion of the set of instantaneous eigenvectors of $H_{\textrm{BdG}}(t)$, which we define as the set \{$\psi_\mu(t)\}$, 
as $\Psi_\mu(t)=\sum_\nu c^{\mu}_{\nu}(t)\psi_\nu(t)$,
where $c^{\mu}_{\nu}(t)\in \mathbb{C}$.

Using the time-dependent BdG equation (TDBdG), we track the evolution of the $c^{\mu}_{\nu}(t)$ coefficients via the following equation of motion:
\begin{equation}
    \begin{split}
    \dot{c}^\mu_{\nu}(t)=-i\left(E(t)-A(t)\right)_{\nu \lambda}c^{\mu}_{\lambda}(t)=-iM_{\nu \lambda}(t)c^{\mu}_{\lambda}(t).
    \end{split} 
\end{equation}
Here, $E_{\mu \nu}(t)=\psi_\mu^{\dag}(t)H_{\textrm{BdG}}(t)\psi_\nu(t)$.
$A_{\mu \nu}(t)=i\psi^{\dag}_\mu(t)\dot{\psi}_\nu(t)$ is the Berry matrix,\cite{Simon83,berry84,wilczek1984,bonderson2011}.
We find
\begin{equation}
    c^{\mu}_{\nu}(t)=\mathcal{T}\exp(-i\int^t_{t_0}M(t')dt')_{\nu \lambda}c^{\mu}_{\lambda}(t_0)\label{eq:transfermat}
\end{equation}
which was derived previously by Cheng \emph{et al.} for $M(t)=M$\,\cite{cheng2011}. 
A varying $M(t)$ was considered by Scheurer and Shnirman, focusing on non-adiabatic transitions\,\cite{scheurer2013}. 

Focusing on the Majorana subspace, we define $\hat{d}^{\dag}_i(t)=\frac{1}{\sqrt{2}}\big(\hat{\gamma}_{2i-1}(t)-i\hat{\gamma}_{2i}(t)\big)$, where $\hat{\gamma}_i$ is an MZM operator.
We map the time instantaneous wavefunctions $\{\psi_\mu(t)\}$ to the instantaneous MZM wavefunctions, $\{\psi^{\gamma}_i(t)\}$, by a change of basis matrix, $W(t)$, such that $\Psi_\mu(t)=c^{\mu}_{\lambda}(t)W^{\dag}_{\lambda j}(t)\psi^{\gamma}_j(t)$.
If we set $\Psi_i(0)=\psi^\gamma_i$, this will now correspond to the time-evolved MZM wavefunction.
Qubit information may then be obtained by analyzing time-dependent Pauli projections $\langle \hat{\vec{\sigma}}(t)\rangle$.  
For a four MZM system, these are given by $(\hat{\sigma}_x,\hat{\sigma}_y,\hat{\sigma}_z)=2i(\hat{\gamma}_3\hat{\gamma}_2,\hat{\gamma}_1\hat{\gamma}_3,\hat{\gamma}_2\hat{\gamma}_1)$. 
For more details 
see the Supplementary Material (SM)\,\cite{supp}.
\begin{figure}[t!]
    \centering
    \includegraphics{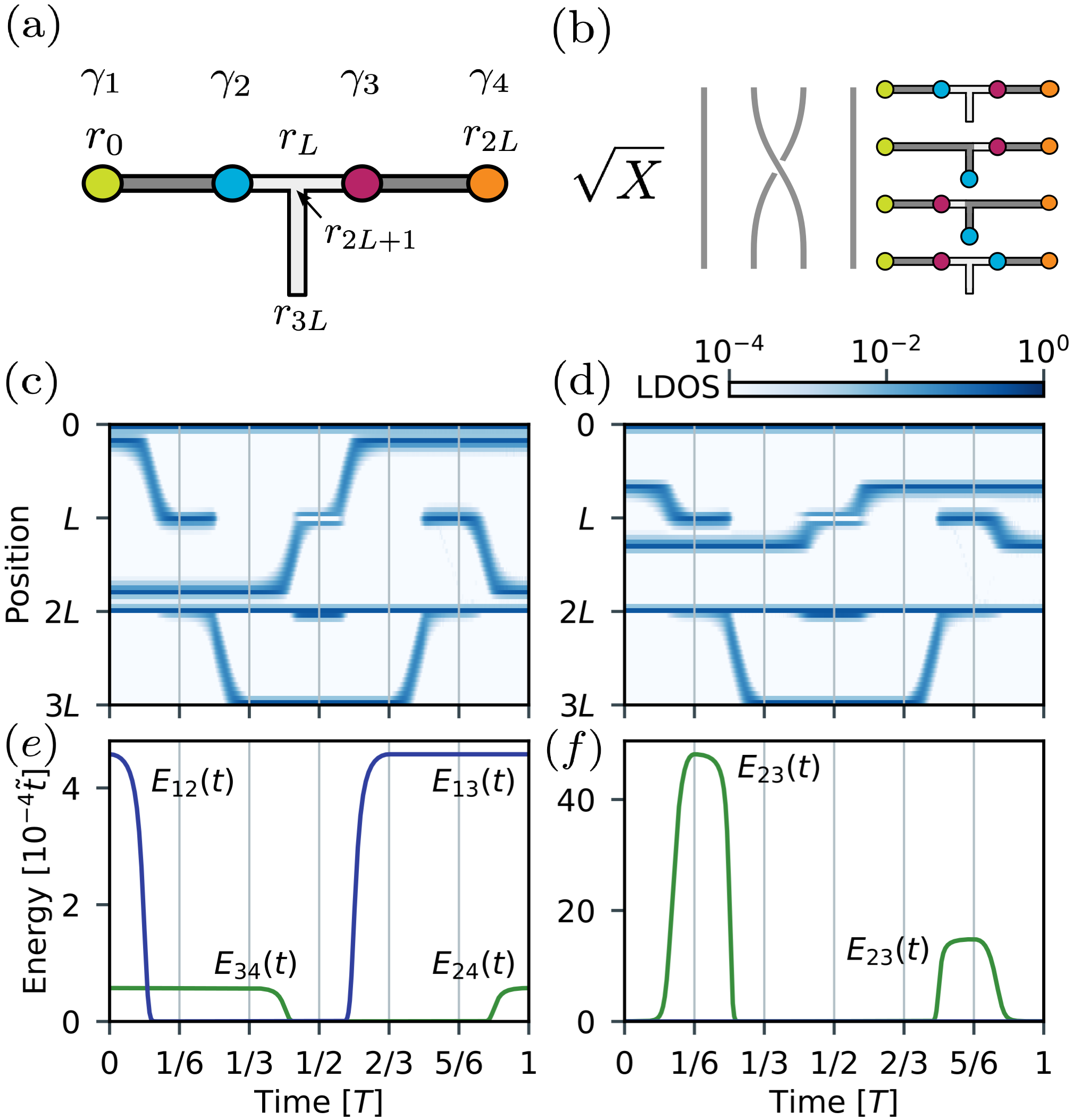}
    \caption{
   (a) Setup of a T-junction with two pairs of MZMs, separated by a trivial region. (b) $\sqrt{\textrm{X}}$ gate braiding protocol. 
   (c) The LDOS of MZM wavefunctions in the energy-splitting  and (d) tunneling regime, respectively. (e, f) The instantaneous energies of the MZMs for both regimes. 
   }
    \label{fig:Fig1}
\end{figure}
{\it Results.---} We consider a trijunction of Kitaev's quantum wire model\,\cite{kitaev2001,alicea2011}, with each leg given by Eq.\,\eqref{eq:MFHam} with  $h_{ab}=-\mu_a(t)\delta_{ab} -\tilde{t} (\delta_{ab+1} + \delta_{a+1b})$ and $\Delta_{ab}=e^{i\phi_p}|\Delta_p| (\delta_{ab+1} - \delta_{b+1a})$.
A graphic of this model is given in Fig.\,\ref{fig:Fig1}\,a.
We will set $|\tilde{t}|=|\Delta_p|=1$, with $\phi_p$=$0$, $\pi$, $\frac{\pi}{2}$ along the left, right and bottom leg, respectively.
%
For a single qubit, information is stored in the two-level state $|\psi\rangle=\textrm{cos}(\frac{\theta}{2})|0\rangle_{\textrm{log}}+e^{i\phi}\textrm{sin}(\frac{\theta}{2})|1\rangle_{\textrm{log}}$, where $\theta\in [0,\pi]$, $\phi\in [0,2\pi]$.  
For $|0\rangle_d$ the unfilled ground state, satisfying $\hat d_\mu|0\rangle_d=0$, we define $|\vec{n}_d\rangle_{\textrm{phys}}=\prod_\nu(d^{\dag}_\nu)^{n_\nu}|0\rangle_d$, where $n_\nu$ is the occupancy of the $\nu$th state. 
We may then translate from the physical to the logical basis (even parity) by $|0\rangle_{\textrm{log}}=|00\rangle_{\textrm{phys}}$, $|1\rangle_{\textrm{log}}=\hat{d}_1^{\dag}\hat{d}_2^{\dag}|0\rangle_d=|11\rangle_{\textrm{phys}}$.
Finally, all results will be found in the adiabatic limit, thus suppressing diabatic transitions over the time-evolution.

{\it X-Gate Oscillations.---} To analyze Majorana hybridization, we perform the X-gate braiding routine, illustrated in Fig.\,\ref{fig:Fig1}\,b\,\cite{alicea2011}, by dynamic control of the chemical potential $\mu_i(t)$ on each site $i$.
The low energy Hamiltonian reads 
$\mathcal{H}_{\textrm{eff}}(t)=\frac{i}{2}\sum_{ij}E_{ij}(t)\hat{\gamma}_i(t)\hat{\gamma}_j(t)$ \cite{bauer2018} with
the energy coupling 
$E_{ij}(t)\sim e^{-r_{ij}(t)/\zeta(t)}$, where $r_{ij}$ the spatial separation between $\gamma_i$ and $\gamma_j$ and $\zeta$ the decay length of the Majorana wavefunction\cite{cheng2009}.
This implies an energy coupling between $\gamma_i$ and $\gamma_j$ that scales with the wavefunction overlap of each MZM. 
The particle hole wavefunctions associated with each energy splitting $E_{ij}(t)$ is given by $\psi^{p/h}_{ij}(t)=\psi^{\gamma}_{i}(t) \mp i\psi^{\gamma}_{j}(t)$, where $p$ $(h)$ labels the particle (hole) state and $\psi^{\gamma}_{i}(t)$ is the wavefunction of $\gamma_i$.
We expand the dynamic wavefunction $\Psi_i(t)$ in the subspace spanned by $\{\psi^p_{12}(t),\psi^h_{12}(t),\psi^p_{34}(t),\psi^h_{34}(t)\}$. 
We may write the coupling matrix, $M(t)$, in the following block matrix form
\begin{equation}
M(t) =\begin{pmatrix}
       M_{12}(t)& M_{23}(t)\\ 
        M^T_{23}(t) & M_{34}(t)
    \end{pmatrix}=M^{(1)}(t)+M^{(2)}(t),
\end{equation}
where $M_{12}(t)=E_{12}(t)\sigma_z$ $+A_{12}(t)\sigma_x$ and $M_{23}(t)=-(\frac{E_{23}(t)}{2}+A_{13}(t))(\sigma_z+i\sigma_y)$.  
Further, $M_{34}(t)$ has the same form as $M_{12}(t)$ up to index labelling, see SM\,\cite{supp}.
$M^{(1)}(t)$ corresponds to the block diagonal terms of $M(t)$ that drive transitions between the particle-hole states of same $\gamma_i$. 
Whereas, $M^{(2)}(t)$ are the off-diagonal contributions that lead to a tunneling between $|\psi^{p/h}_{12}(t)\rangle$ and $|\psi^{p/h}_{34}(t)\rangle$. 
The case where $M^{(2)}(t)$ ($M^{(1)}(t)$) terms are suppressed, we will denote as the \emph{energy-splitting} (\emph{tunneling}) regime. 

\begin{figure}[t!]
    \centering
    \includegraphics{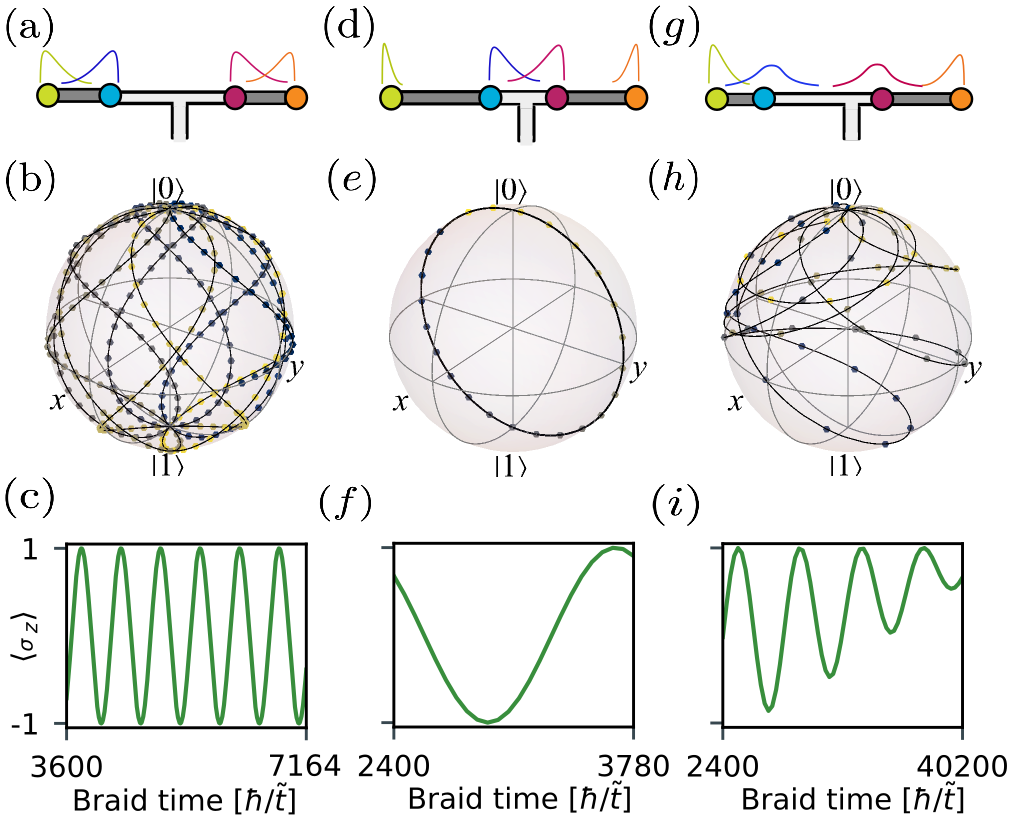}
    \caption{X-gate oscillations due to energy-splitting (first column), tunneling (second column) and combined-case regimes (third column). (a, d, g) Illustration of wavefunction overlaps.
    (b, e, h) 
    The final state outcomes, $|\Psi(T)\rangle$, as a function of total braid time, $T$, for the X Gate, with analytical final state trajectories (solid line) plotted against many-body simulations (dots). (c, f, i) The Pauli-Z projection, $\langle \hat{\sigma}_z(T)\rangle$, as a function of $T$.
    %
    The parameter set for each case ($\mu_{\textrm{topo}}$, $\mu_{\textrm{triv}}$, $r_{12}$, $r_{34}$, $L$)=$(0.8,10,4a_0,12a_0,24a_0)$, $(0.0,5.2,3a_0,10a_0,15a_0)$, $(0.4,4.5,4a_0,10a_0,20a_0)$ and initial state $|\psi\rangle=|0\rangle_{\rm log}$.}
    \label{fig:Fig2}
    %
\end{figure}

Initializing the state $|\psi\rangle=|0\rangle_{\textrm{log}}$, we simulate the braid for different total braid times $T$\,\cite{mascot2023}. 
We first analyze the energy-splitting regime, with dominant hybridization through the topological region. 
The overlap between the Majorana wavefunctions can be seen in the local density of states (LDOS), Fig.\,\ref{fig:Fig1}\,c. 
Initially, the Majorana wavefunction overlap is between $\psi^{\gamma}_1$ $(\psi^{\gamma}_3)$ and $\psi^{\gamma}_2$ $(\psi^{\gamma}_4)$, as depicted in Fig.\,\ref{fig:Fig2}\,a. This induces an energy splitting $E_{12}(t)$ ($E_{34}(t)$), as seen in Fig.\,\ref{fig:Fig1}\,e.
However, after the swap across the junction at $r_{L}$, $\psi^{\gamma}_1$ $(\psi^{\gamma}_2)$ now overlaps with $\psi^{\gamma}_3$ $(\psi^{\gamma}_4)$, leading to $E_{13}(t)$ and $E_{24}(t)$ contributions.
For the Berry-matrix transitions, $A_{12}(t)$, $A_{34}(t)\sim v(t)e^{-r_{ij}/\zeta(t)}$, where $v(t)$ is a time-dependent function that scales with the speed of the process.
In the adiabatic limit, these get suppressed, along with $M^{(2)}(t)$, as $v(t)\to 0$\,\cite{cheng2011, scheurer2013}. 
 We now solve analytically for the expectation values of the Pauli operators: 
\begin{align}
             &\langle\hat{\sigma}_x(T)\rangle=-\textrm{cos}\left(\bar{E_2}T\right)\textrm{sin}\left((\bar{E}_1+\bar{E_2})T\right)\ ,
   \\[5pt]
            &\langle\hat{\sigma}_y(T)\rangle=\textrm{sin}\left(\bar{E_2}T\right)\textrm{sin}\left((\bar{E}_1+\bar{E_2})T\right)\ ,
    \\[5pt]
             &\langle \hat{\sigma}_z(T)\rangle=- \textrm{cos}\left((\bar{E}_{1}+\bar{E}_{2})T\right)\ .
\end{align}
We note that a successful $X$ braid would correspond to $\langle \hat{\sigma}_z(T)\rangle=-1$.
Here, we define $\bar{E}_{ij}=\frac{1}{T}\int^T_{0}E_{ij}(t')dt'$ to be the average coupling energy between $\gamma_i$ and $\gamma_j$. 
Further, we set $\bar{E}_1=\bar{E}_{12}+\bar{E}_{34}$ and $\bar{E}_2=\bar{E}_{13}+\bar{E}_{24}$.
These results are compared to many-body simulations in Fig.\,\ref{fig:Fig2}\,b and c.
Notably, the $\langle \hat{\sigma}_z(T)\rangle$ projection, an equivalent measure of the transition probability, $|\langle 1|0(T)\rangle_{\rm log}|^2$, oscillates as a function of total braid time.
We further highlight that in the odd parity subspace, $\bar{E}_{1/2}=\bar{E}_{1(2/3)}-\bar{E}_{(3/2)4}$,
providing a potential comparative advantage for suppressing qubit error (see SM\,\cite{supp}).
This yields a striking result: we can predict the final state and thus also the fidelity solely based on the instantaneous energies, without the need of a dynamical calculation.


Next we study braiding in the tunneling regime, 
where $E_{23}(t)$ is the dominant coupling. 
To this end, we initialize the system as 
depicted in Fig.\, \ref{fig:Fig1}\,d, where Majorana separation $r_{23}$ reaches a minimal 
point crossing the junction and $E_{23}(t)$ reaches a maximum, as shown in Fig.\,\ref{fig:Fig1}\,f.
In this case, the final state projections are given by 
\begin{align}
    &\langle \hat{\sigma}_x(t)\rangle=0\ ,\\[5pt]
    &\langle \hat{\sigma}_y(t)\rangle=-\textrm{sin}(2\bar{E}_{23}T+4\beta)\ ,\\[5pt]
    &\langle \hat{\sigma}_z(t)\rangle=-\textrm{cos}(2\bar{E}_{23}T+4\beta)
\end{align}
where $\beta=\int^T_0A_{23}(t')dt'$ is a path dependent quantity, and is independent of $T$.
Unlike the energy-splitting case, the final state is restricted to the $yz$-plane of the Bloch-sphere, as shown in Fig.\,\ref{fig:Fig2}\,e. 
However, we similarly find a period of oscillation scales with the average energy splitting, $\tau=\pi/\bar{E}_{23}$.


Finally, we consider the case where both the energy-splitting and tunneling contribute to $M(t)$. 
As shown in Fig.\,\ref{fig:Fig2}\,g, we initialize the system with dominant $\gamma_1$-$\gamma_2$ wavefunction overlap, inducing an $E_{12}(t)$ coupling. However, as $\gamma_2$ approaches the junction, the key hybridization channel becomes $E_{23}(t)$ as $\gamma_2-\gamma_3$ overlap starts to dominate.
Much like the energy splitting case, 
Fig.\,\ref{fig:Fig2}\,h resembles a web-like structure around the Bloch sphere.
However, as can be seen in Fig.\,\ref{fig:Fig2}\,i, the $\langle \hat{\sigma}_z(T) \rangle$ projection fails to reach $-1$ at the minimum, constituting a failed braid at all braid times $T$ calculated. 
  This makes it a difficult, however, possible, case to tune $T$ for a correct braiding outcome, with $\langle \hat{\sigma}_z(T)\rangle =\frac{1}{2}(1-\cos{((\bar{E}_{12}+\bar{E}_{13})T)}-\cos{(4\beta+2\bar{E}_{23} T)}-\cos{(4\beta+2\bar{E}_{23} T)}\cos{((\bar{E}_{12}+\bar{E}_{13})T)})$

These results have immediate implications for the speed of a successful braid.
For the lowest lying bulk state energy, $ E_{\rm {bulk}}$, and total hybridization energy, $E_{\rm {hyb}}$, it is well known that $E_{\rm {bulk}}^{-1}\ll T \ll E_{\rm {hyb}}^{-1}$\,\cite{Thouless1991,nayak2008,Brouwer2011,sanno_it_2021}. 
For $E_{\rm {hyb}}=\bar{E}_1+\bar{E}_2$, $2\bar{E}_{23}$, in the energy splitting and tunneling case respectively, we quantify this by asserting $|\langle 1_{\rm log}|0_{\rm log}(t)\rangle|^2=\rm {cos}^2(\frac{E_{\rm {hyb}}T}{2})>0.99$. to satisfy Pauli error correction thresholds \cite{Raussendorf2007,Stace2009}. 
This provides a sensitive,  gate-dependent  bound for $T$ (see SM\,\cite{supp}).
\begin{figure}[t!]
    \centering
    \includegraphics{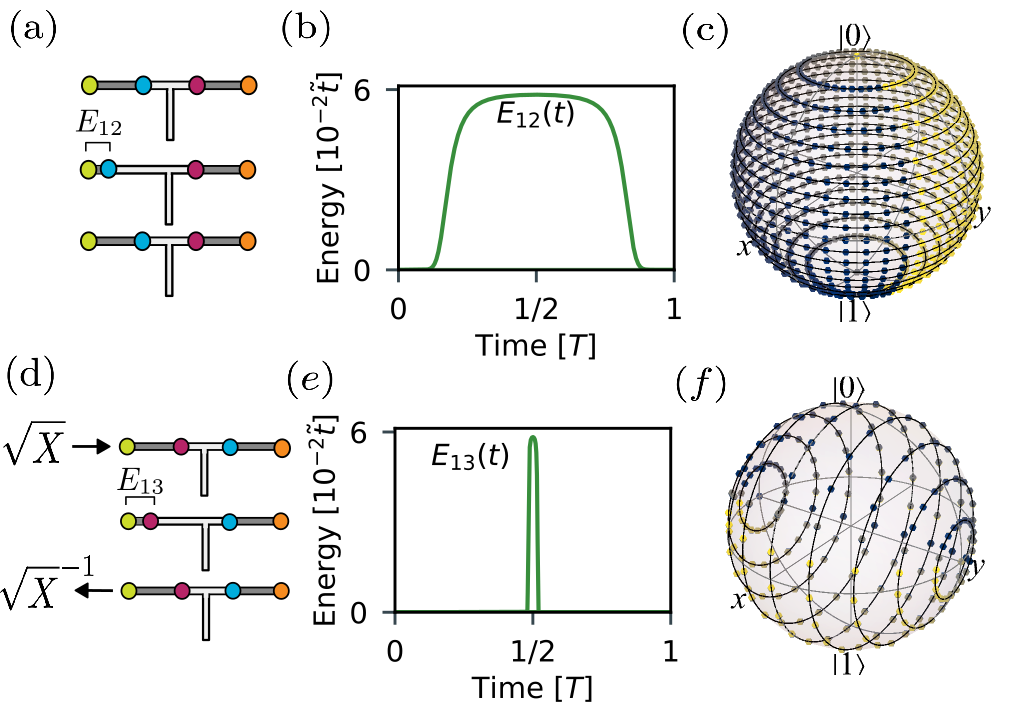}
    \caption{Single-Qubit gate simulations: (a) Schematics of $R_z(\theta)$-gate protocol. (b) Real-time instantaneous MZM energy during phase gate protocol. (c) Simulation of phase gate for different states $|\psi\rangle$ with varying $\theta \in [0,\pi]$ and fixed $\phi=0$. (d) Schematics of $R_y(\theta)$ gate protocol. (e) Real-time instantaneous MZM energy during $R_y(\theta)$ procedure. (f) Simulation of $R_y(\theta)$ gate with $\theta=\pi$ and varying $\phi\in [0,2\pi]$. 
    In (c), (f) analytical final state trajectories (solid line) plotted against many-body simulations (dots).
    Parameters for both simulations: 
    ($\mu_{\textrm{topo}}$, $\mu_{\textrm{triv}}$, $r_{12}$, $r_{34}$, $L$)=$(0.35,10,10a_0,10a_0,20)$ with $\Delta r=8a_0$, $a_0$ being the lattice spacing.}
    \label{fig:Fig4}
\end{figure}
{\it Qubit gates I: Phase gate.---} We now utilize these findings to implement and simulate single-qubit gates using hybridization, essential for the successful operation of a universal quantum computer.

We begin with the $R_z(\varphi)$ gate, encompassing an arbitrary rotation by $\varphi$ about the $z$-axis (i.e., $|1\rangle_{\textrm{log}} \to e^{i\varphi}|1\rangle_{\textrm{log}}$).
For $\varphi=\pi/4$, this is the magic or T-gate, necessary for a full mapping onto the Bloch-sphere and unattainable via usual braiding procedures\,\cite{bravyi2002,Bonderson2010}. 
We induce an energy splitting by moving $\gamma_{2}$ towards $\gamma_1$ by a displacement $\Delta r$ (Fig.\,\ref{fig:Fig4}\,a). 
Afterwards, we move them back to their initial positions with negligible hybridization.
The dynamic energy $E_{12}(t)$ 
 is given in Fig.\,\ref{fig:Fig4}\,b;
the phase accumulation again depends on the average energy splitting, with $\varphi=\bar{E}_{12}T$. 
This is displayed in Fig.\,\ref{fig:Fig4}\,b, where a state with arbitrary initial $\hat{\sigma}_z$ projection, $\langle\hat{\sigma}_z\rangle=\textrm{cos}(\frac{\theta}{2})$, will rotate about the z-axis, controlled by simply changing the protocol time $T$.

We extend this to an arbitrary rotation about the y-axis, an $R_y(\varphi)$ gate. 
This is achieved by the routine $R_y(\varphi)=\sqrt{\textrm{X}}^{-1}R_z(\varphi)\sqrt{\textrm{X}}$, where the wavefunction overlap between $\gamma_1$ and $\gamma_3$ dictates the phase accumulation. 
The routine for this is depicted in Fig.\,\ref{fig:Fig4}\,d, with the energy coupling $E_{13}(t)$ gained after the first $\sqrt{\textrm{X}}$ gate given in Fig.\,\ref{fig:Fig4}\,e. 
As demonstrated in Fig. \ref{fig:Fig4}\,f, for an arbitrary state, the $\langle \hat{\sigma}_y\rangle$ remains fixed as the state rotates about the $y$-axis with phase $\varphi=\bar{E}_{13}T$, clockwise from the positive $z$ axis. 
This constitutes enough for an arbitrary mapping of any state $|\psi\rangle$ onto the Bloch-sphere, as we may decompose an arbitrary rotation about the Bloch-sphere, into a set of Euler rotations\,\cite{Schmidt2013}. 

%


\begin{figure}[t!]
    \centering
    \includegraphics{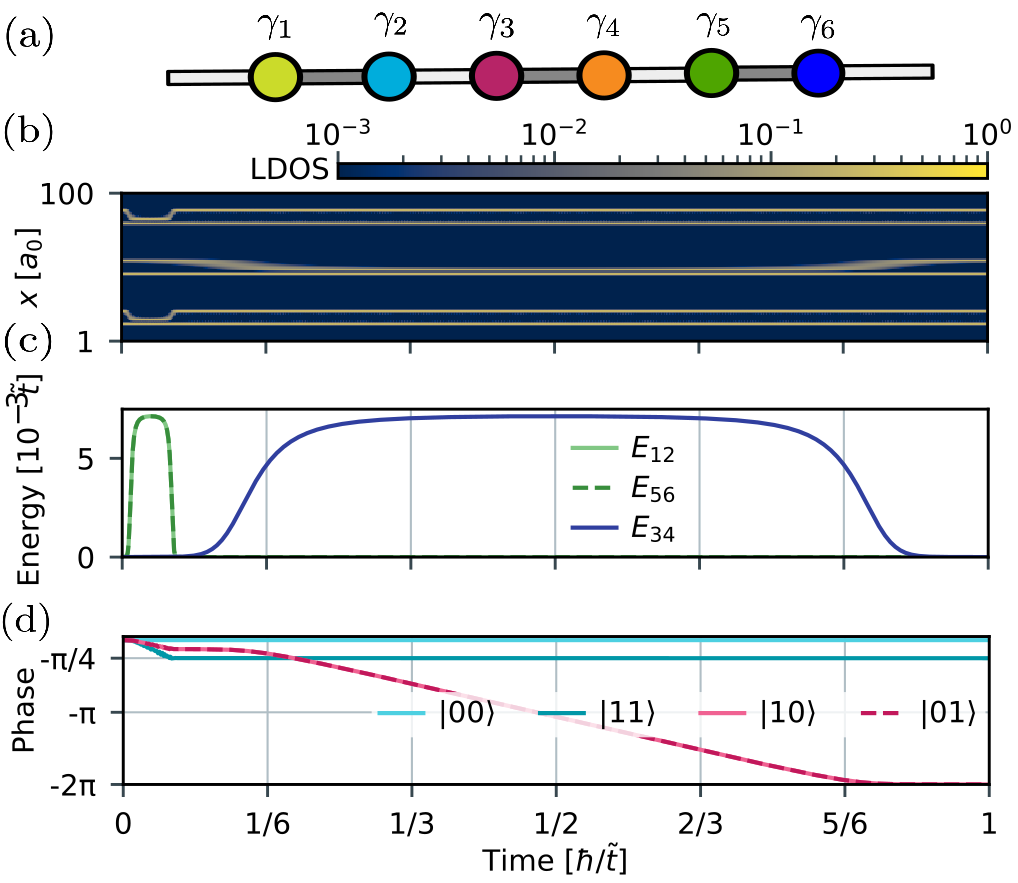}
    \caption{CT-gate via hybridization. (a) Kitaev chain with three topological segments and three pairs of MZMs. (b) LDOS over time $t$ in the ground-state manifold. (c) Plot of instantaneous energies as a function of real time $t$ for each Majorana state. (d) Plot of the phase evolution as a function of time for each basis state in the ground-state subspace. Parameters used: ($\mu_{\textrm{topo}}$, $\mu_{\textrm{triv}})$ =$(0.4,12)$ with $r_{12}$=$r_{34}$=$r_{56}$=$10a_0$ and $r_{23}$=$r_{45}$=$20a_0$. Chain length $L=100a_0$.}
    \label{fig:Fig5}
\end{figure}

{\it Qubit gates II: Controlled-Phase gate.---} Finally, we extend phase-gate results to implement a two-qubit controlled phase gate, purely through hybridization of MZMs. 
Since this requires no braids, we 
simply consider a chain, 
 (Fig.\,\ref{fig:Fig5}\,a).
Using the dense qubit design\,\cite{dassarma-15npjqi15001} we initialize the system with three pairs of  MZMs instead of two, with $|\psi\rangle=c_{1}|000\rangle_{\textrm{phys}} +c_{2}|110\rangle_{\textrm{phys}}+c_{3}|101\rangle_{\textrm{phys}}+c_{4}|011\rangle_{\textrm{phys}}$.
For an arbitrary phase $\varphi$, a successful controlled phase gate is achieved by mapping $|101\rangle_{\textrm{phys}} 
\to e^{i\varphi}|101\rangle_{\textrm{phys}}$, 
leaving the other states unchanged.
As shown in Fig.\,\ref{fig:Fig5}\,(b, c), this may be achieved 
by inducing an energy splitting through the topological region by bringing $\gamma_2$ ($\gamma_6$) toward $\gamma_1$ ($\gamma_5$), inducing an average energy splitting $\bar{E}_{12}$ ($\bar{E}_{56}$). 
A state $|nlm\rangle_{\textrm{phys}}$ will subsequently map to $e^{-i(n\varphi_{12}+l\varphi_{34}+m\varphi_{56})}|nlm\rangle_{\textrm{phys}}$, with $n,l,m\in \{0,1\}$ indexing the occupation of the single-particle states $|\psi^p_{ij}(t)\rangle$, and $\varphi_{ij}=-\bar{E}_{ij}T_{ij}$ the accumulated phase of that state, with $T_{ij}$ the set hybridization time between $\gamma_i$ and $\gamma_j$.
For an arbitrary choice of $\varphi$, we set $\bar{E}_{12}T_{12}=\bar{E}_{56}T_{56}=\varphi/2$, leading to a total accumulation of $\varphi$ on the state $|101\rangle_{\textrm{phys}}$, as necessary. 
To avoid a phase accumulation on $|110\rangle_{\textrm{phys}}$, $|011\rangle_{\textrm{phys}}$, we set $\bar{E}_{34}T_{34}=2\pi-\varphi$.
For fixed $\bar{E}_{12}=\bar{E}_{34}=\bar{E}_{56}=4.35\times 10^{-3}$, we set the total braid time $T_{12}=T_{56}=40.60\hbar$ and $T_{34}=609.06\hbar$.
As shown in Fig.\,\ref{fig:Fig5} (d), the phase accumulation for each state is simulated in real time.
For the $|101\rangle_{\textrm{phys}}$ state plateau at $t=T_{12}$ at $-\pi/4$, while $|110\rangle$ and $|011\rangle$ converge to $-2\pi$, as required for a controlled-T (CT) or controlled magic gate. 
This constitutes the successful simulation of a CT-gate,
purely through hybridization, on a Majorana-based platform, unattainable through braiding processes.

{\it Outlook.---} 
We have provided an accessible yet accurate pathway to analyzing the dynamic evolution of MZMs in the presence of hybridization processes, vital for the operation of a fault-tolerant topological quantum computer. 
While we demonstrate this in the context of four and six-MZM systems, showing how to track braiding errors and how to
institute single-qubit gates forbidden in the Ising anyon model, we stress that this method is scalable to larger MZM systems. 
%
%
Further, this method applies not only to the Kitaev chain, but any model including superconductor-semiconductor hybrid systems \cite{Karzig2019}, magnetic superconducting hybrid systems \cite{nadjperge13,schneider21,bedow2024} and skyrmion spin texture platforms \cite{Konakanchi2023}.
%

Important open questions are related to generic errors of braiding with MZMs such as decoherence sources, higher energy states, electrostatic environment and gate noise. 
We believe that any source of error that will not increase the hybridization energy $\bar{E}_{ij}$, as explicated in this paper, will lead to diabatic effects, corresponding to the lower bound for the speed limit of MZMs discussed previously\,\cite{nayak-08rmp1083,Thouless1991,Brouwer2011,sanno_it_2021}. 
The interplay of these two effects can be non-trivial, as recently demonstrated for braiding in the presence of static disorder\,\cite{Peeters2024}. 
Quantifying the effect and dynamics of various types of diabatic error
remains an open challenge\,\cite{cheng2011,scheurer2013,knapp_the-nature_2016,karzig_boosting_2013,conlon-19prb134307,sekania2017}. 
We stress, however, that any type of error contributing to an increase in MZM hybridization is quantified by the results of this paper.

\begin{acknowledgments}
S.R.\ acknowledges support from the Australian Research Council through Grants No.\ DP200101118 and DP240100168.
This research was undertaken using resources from the National Computational Infrastructure (NCI Australia), an NCRIS enabled capability supported by the Australian Government.
\end{acknowledgments}

\bibliography{Hybrid_MZM}

\end{document}


\title{Supplemental Material:\\[5pt]
Characterizing Dynamic Majorana Hybridization\\
for Universal Quantum Computing}

\author{Themba Hodge}
\author{Eric Mascot}

\author{Dan Crawford}
\affiliation{School of Physics, University of Melbourne, Parkville, VIC 3010, Australia}
\author{Stephan Rachel}
\affiliation{School of Physics, University of Melbourne, Parkville, VIC 3010, Australia}
\noaffiliation

\date{\today}

\maketitle

%

\tableofcontents

\section{Model}
We utilise the one-dimensional topological superconductor, the Kitaev chain\,\cite{kitaev2001}, to study the dynamics of emergent MZMs. 
The Hamiltonian for this system is given by
\begin{align}
H = -\mu \sum_a c_a^\dagger c_a^\pd
- \tilde{t} \sum_a \left(
    c_a^\dagger c_{a+1}^\pd + \hc
\right)
+ \sum_a \left(
    \Delta_p c_a^\dagger c_{a+1}^\dagger + \hc
\right),
\end{align}
where $\mu$ is the chemical potential, $\tilde{t}$ is the hopping amplitude, and $\Delta_p = |\Delta_p| e^{i\phi}$ is a phase dependent $p$-wave pairing amplitude.\\ 

For the purpose of braiding, we extend this one-dimensional geometry to a quasi two-dimensional geometry by means of a trijunction or T-junction\,\cite{alicea2011, mascot2023}. 
This consists of three Kitaev chains of length $L$, connected at an central sight, known as the junction $r_0$. 
The Hamiltonian for the T-junction is given by
\begin{figure}[t!]
    \centering
    \includegraphics{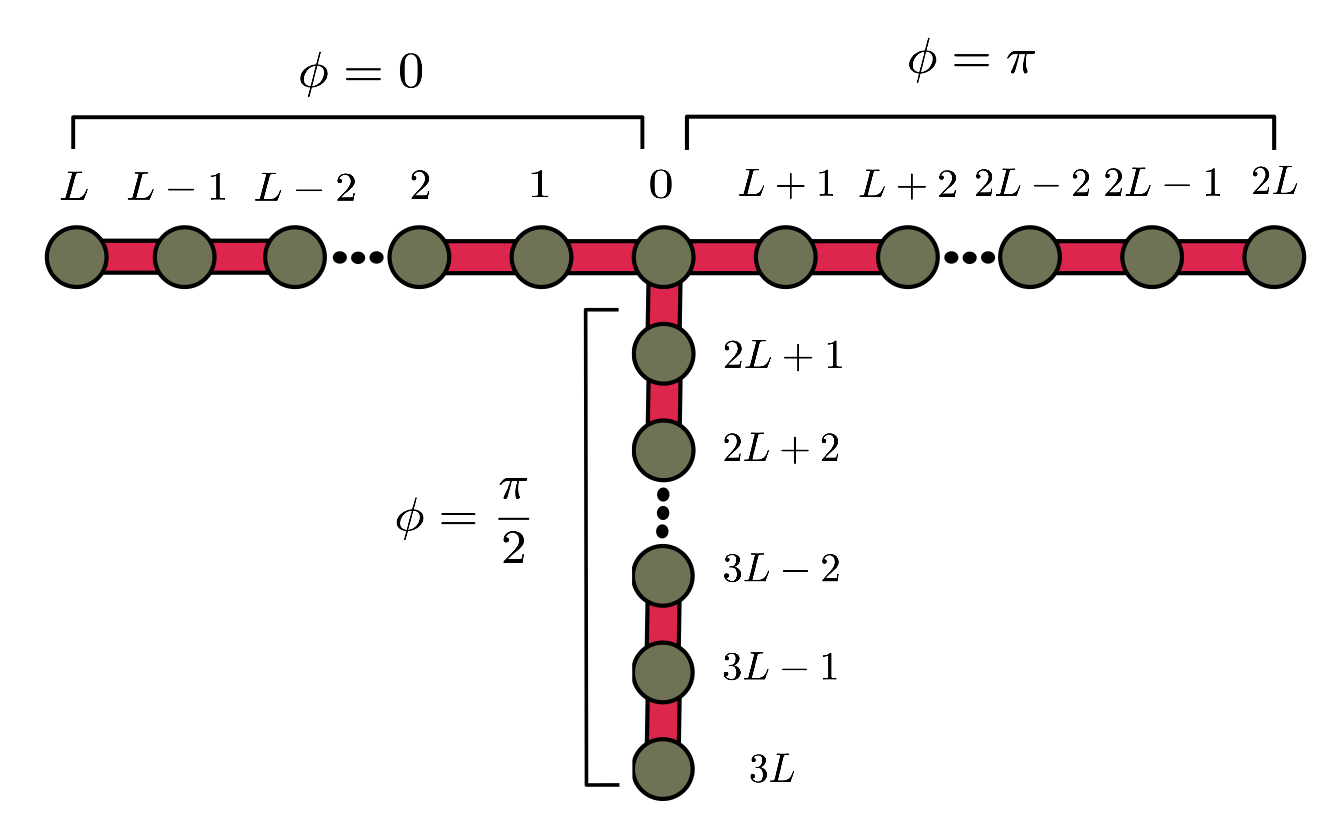}
    \caption{Lattice geometry of a T-junction. The lattice site indices are labeled above the site along with the superconducting phases $\phi$ along each leg.}
    \label{fig:Fig1sup}
\end{figure}
\begin{align}
\mathcal{H} &=
- \sum_{a=0}^{3L} \mu_x c_a^\dagger c_a^\pd
+ \sum_{n=0}^2 \left(
    -\tilde{t}\, c_0^\dagger c_{nL+1}^\pd
    + \Delta_p^{(n)} c_0^\dagger c_{nL+1}^\dagger
    + \hc
\right) \nonumber \\
&\hspace{1em}
+ \sum_{a=0}^2 \,\sum_{a=nL+1}^{(n+1)L} \left(
    - \tilde{t}\, c_a^\dagger c_{a+1}^\pd
    + \Delta_p^{(n)} c_a^\dagger c_{a+1}^\dagger + \hc
\right).
\end{align}

Geometry and site-labeling of the T-junction is shown in Fig.\,\ref{fig:Fig1sup}.
We choose the superconducting phases as $\Delta_p^{(0)} = |\Delta_p|$, $\Delta_p^{(1)} = -|\Delta_p|$, and $\Delta_p^{(2)} = i|\Delta_p|$ corresponding to phases $\phi_0=0$, $\phi_1=\pi$ and $\phi_2=\pi/2$, and aligning with the angle of the leg from the central junction.
The topological regime is realized 
for $-2\tilde{t}<\mu_{\textrm{topo}}<2\tilde{t}$, with MZMs emergent on the chain ends.

Lastly, we move Majorana zero-modes (MZMs) by means of a time-dependent chemical potential along each site, which will adjust the boundary of the topological region of the system. 

We do this via a smooth-step polynomial function, $r$, which will ramp the chemical potential along each leg, during a step, between $\mu_{\textrm{triv}}$ and $\mu_{\textrm{topo}}$. 
For $0 \leq q \leq 1$, where $q=0$ denotes the beginning of the step, and $q=1$ the end of the step, this function is given by

\begin{align}
    s(q) = \begin{cases}
    0 & q \leq 0, \\
    q^2 (3-2q) & 0 \leq q \leq 1, \\
    1 & 1 \leq q
    \end{cases}.
\end{align}

As such, we may encode the time-dependent chemical potential at each lattice site along a leg $r=i$ as 

\begin{equation}
    \mu_a (t) = \mu_{\rm triv} + (\mu_{\rm topo} - \mu_{\rm triv}) \,\,s\left( \frac{t}{\tau}(1 + \alpha (N-1)) - \alpha r \right), 
\end{equation}

where $\alpha$ is a \emph{delay coefficient} that scales the delay between the start of the ramp along each site on the leg\,\cite{mascot2023}. 
For the entirety of this paper $\alpha$ is set to $0.025$. 


\section{Equation of Motion}
Consider a set of time-dependent Boguliubov quasiparticles $\{a(t), a^{\dag}(t)\}$, where  $\hat{a}_n(t)=\mathcal{S}(t,t_0)\hat{a}_n(t_0)\mathcal{S}^{\dag}(t,t_0)$. 
We first note that $\hat{a}_n$ may be connected to the bare fermionic basis ${\hat{c},\hat{c}^{\dag}}$ by the Boguliubov coherence factors, i.e., 
\begin{equation}
    \begin{pmatrix}
    \hat{c} \\ 
    \hat{c}^{\dag} \end{pmatrix}=\begin{pmatrix}
    U & V^* \\ 
    V  & U^*
    \end{pmatrix}\begin{pmatrix} \hat{a}\\ 
    \hat{a}^{\dag} \end{pmatrix}.
\end{equation}
This corresponds to a Boguliubov transformation, with $U$ and $V$ are $n\times n$ dimensional matrices and $U^*$ and $V^*$ their conjugates. 
Similarly, the time-evolved quasi-particles may be connected to the bare fermionic basis by  
\begin{equation}
    \begin{pmatrix}
    \hat{c} \\ 
    \hat{c}^{\dag} \end{pmatrix}=\begin{pmatrix}
    U(t) & V(t)^* \\ 
    V(t)  & U(t)^*
    \end{pmatrix}
    \begin{pmatrix} \hat{a}(t)\\ 
    \hat{a}^{\dag}(t) \end{pmatrix} \label{eq:TevBog}
\end{equation}
Using the time-dependent Bogoliubov--de Gennes (TDBdG) equation\,\cite{kummel1969,cheng2011}, $U(t)$ and $V(t)$ may be found by solving the following equation of motion:
\begin{align}
&i\partial_t\begin{pmatrix}
    U(t) & V(t)^* \\ 
    V(t)  & U(t)^*
    \end{pmatrix}=H_{\textrm{BdG}}(t)\begin{pmatrix}
    U(t) & V(t)^* \\ 
    V(t)  & U(t)^*
    \end{pmatrix}\\[7pt] 
    \implies &
    \begin{pmatrix}
    U(t) & V(t)^* \\ 
    V(t)  & U(t)^*
    \end{pmatrix}=
    \mathcal{T}\textrm{exp}\left(-i\int^t_0H_{\textrm{BdG}}(t')dt'\right)\begin{pmatrix}
    U & V^* \\ 
    V  & U^*
    \end{pmatrix} \label{eq:TDBdG}
\end{align}
where $U(t=0)=U$ and $V(t=0)=V$. 
Here, $\mathcal{T}$ corresponds to the time-ordering operator and $H_{\rm BdG}$ is the mean-field Boguliubov--de Gennes (BdG) Hamiltonian of the system. 
%
Using this, we adopt the same strategy as Cheng {\it et al.}\,\cite{cheng2011}. 
For $\hat{a}(t)=\hat{d}(t)$, where $\hat{d}(t)$ are the superconducting quasiparticles, we may set $(U(t) \; V(t))^T=(\Psi_1(t) \; \Psi_2(t) \; ... \Psi_N(t))$, where $\Psi_\mu(t)$ is the dynamic wavefunction of the $i$th quasiparticle. 
As $\Psi_\mu(t)$ remains a Boguliubov wavefunction, we expand our dynamic wavefunction $\Psi_i(t)$ as a superposition of our instantaneous eigenfunction basis $\{\psi_\mu(t)\}$ , i.e., $\Psi_\mu(t)=\sum_\nu c^{\mu}_\nu(t)\psi_\nu(t)$. 
 Here, the $c^{\mu}_\nu(t)$ are complex coefficients and $\{\psi_\mu(t)\}$ correspond to the eigenbasis of $H_{\textrm{BdG}}(t)$. 
 We may take a time-derivative as
 
%
\begin{equation}
    i\frac{d}{dt}\left(\sum_\nu c^{\mu}_\nu(t)\psi_\nu(t)\right)=\sum_\nu \dot{c}^{\mu}_\nu(t)\psi_\nu(t)+c^{\mu}_\nu(t)\dot{\psi}_\nu(t)=\sum_\nu (H_{\textrm{BdG}}(t))_{\mu\nu}c^\mu_{\nu}(t)\psi_\nu(t)
\end{equation}

leading to the equation of motion
\begin{equation}
   \dot{c}^\mu_\nu(t)=-i\left(E(t)-A(t)\right)_{\nu \lambda}c^\mu_\lambda(t)\ . \label{eq:EqofMot} 
\end{equation}
where $E_{\mu \nu}(t)=\langle \psi_\mu(t)|H_{\textrm{BdG}}(t)|\psi_\nu(t)\rangle$ and $A_{\mu \nu}(t)=i\langle\psi_\mu(t)|\dot{\psi}_\nu(t)\rangle$ is the non-Abelian Berry matrix\, \cite{wilczek1984}. 
This encompasses our equation of motion for the $\{c^\mu_\nu(t)\}$ in the ground state manifold, which was derived in \cite{cheng2011}. 

\section{V-Matrix}
Next we solve for the time evolved Majorana operators by a change of basis. 
Consider a system made-up of $N$ MZMs. 
We restrict our interactions to the low energy manifold of the Hilbert space. 
This is covered not only by the physical MZMs, $\{|\psi^{p}_{1}\rangle,...,|\psi^{p}_{N}\rangle\}$, but by their particle-hole conjugates $\{|\psi^{h}_{1}\rangle,...,|\psi^{h}_{N}\rangle\}$. 
These satisfy $|\psi^{p}_{\mu}\rangle=C|\psi^{h}_\mu\rangle$, where $C$ is the charge conjugation operator $C=\tau_xK$.


This restricts our attention to a $2N$-dimensional subspace. Returning to \eqref{eq:EqofMot}, each of the coefficients $\{c^{\mu}_\nu(t)\}$ may be solved for from an initial configuration by taking a time-ordered matrix exponential
\begin{equation}
    c^{\mu}_\nu(t)=\left( \mathcal{T}\textrm{exp}\left(-i\int^t_{t_0}\left(E(t')-A(t')\right)dt'\right) \right)_{\nu \lambda}c^{\mu}_\lambda(t_0)=I_{\nu \lambda}(t,t_0)c^{\mu}_\lambda(t_0)\label{eq: qpart-evol}
\end{equation}
where $c^\mu_\nu(0)=\delta_{\mu \nu}$ corresponds to our initialisation, setting $\Psi_\mu(0)=\psi_\mu(0)$. \\ 


Suppose the dynamical Majorana wavefunctions are given by the expansion 
\begin{equation}
    \Psi_\mu(t)=\sum_{n}b^\mu_j(t)\psi^{\gamma}_j(t), \label{eq:Majadiabatic}
\end{equation}
the coefficient sets $\{b^\mu_j(t)\}$ and $\{c^\mu_\nu(t)\}$ are connected by the change of basis matrix  $W(t)$ which  maps $\{\hat{d}(t)\} \to \{\hat{\gamma}(t)\}$:
\begin{equation}
    b^\mu_j(t)=W^{\dag}_{\nu j}(t)c^{i}_\nu(t).
\end{equation}
This allows us to track the evolution of our MZM wavefunction coefficients $\{b^i_j(t)\}$. Returning to \eqref{eq: qpart-evol}, we obtain 
\begin{equation}\label{eq:ScheurerShnirman}
    b^\mu_j(t)=\left(W^*(t)I(t,t_0)W^T(t_0)\right)_{jk}b^\mu_k(t_0)
\end{equation}
which will encode the evolution of our quasiparticle coefficients due to dynamic transitions between states. 
We note that \eqref{eq:ScheurerShnirman}, when formulated in the quasi-particle basis, corresponds to a formula derived previously by Scheurer and Shnirman\,\cite{scheurer2013}.
We now plug \eqref{eq:ScheurerShnirman} into \eqref{eq:Majadiabatic} to find 
\begin{align}
    \begin{split}
    |\Psi_{\mu}(t)\rangle&=\sum_{j}b^{\mu}_{j}(t) \psi_{j}^{\gamma}(t) \\
                &=\sum_{k}\sum_{j}b^{\mu}_{j}(t)B_{kj}(t,t_0)\psi_{k}^{\gamma}(t_0)\\
                &=\sum_k V_{\mu k}(t,t_0)\psi_{k}^{\gamma}(t_0),
    \end{split}
\end{align}
where
\begin{equation}
    V_{\mu j}(t,t_0)=\left(W^*(t)I(t,t_0)W^T(t_0)\vec{b}(t_0)\right)^T_{\mu k}B^T_{kj}(t,t_0) \label{eq:Vmat}
\end{equation}
and $\vec{b}(t)=(b^\mu_1(t), b^\mu_2(t), b^\mu_3(t), b^\mu_4(t))$ is a matrix containing the expansion coefficients in the Majorana basis. 
$B_{ij}(t,t_0)=(\psi_{i}^{\gamma})^\dag(t_0)\psi_{j}^{\gamma}(t)$ is the mapping between the $\{ \psi^{\gamma}(t_0)\}$ and the $\{ \psi^{\gamma}(t)\}$ eigenbases. 
We now set the initialization to be $b^i_j(0)=\delta_{ij}$. Here, the dynamic wavefunction will correspond to that of a time-evolved MZM, $\Psi^\gamma_i(t)$, rather than the time-evolved quasiparticle wavefunction, $\Psi_\mu(t)$, considered previously. 
 
\section{Single-Qubit Projections onto Bloch-Sphere}
We may now connect the time-evolved Majorana operators $\{\hat{\gamma}_i(t)\}$ to the initial time Majorana operators $\{\hat{\gamma}_i\}$. 
Using the dynamic wavefunctions $\{\Psi^{\gamma}_i(t)\}$, and setting $t_0=0$, we find that
\begin{align}
    \hat{\gamma}_i(t)=(\Psi_i^{\gamma}(t))^{\dag}\begin{pmatrix}
        \hat{c}\\ 
        \hat{c}^{\dagger}
    \end{pmatrix}=\sum_j V^*_{ij}(t)(\psi_j^{\gamma})^{\dag}\begin{pmatrix}
        \hat{c}\\ 
        \hat{c}^{\dagger}
    \end{pmatrix}=\sum_j V^*_{ij}(t)\hat{\gamma}_j.
\end{align} 
We note, however, that for a system with $2\rm{N}$ MZMs, $V(t)\in \textrm{SO}(2\rm{N})$ and therefore $V^*(t)=V(t)$ \cite{cheng2011}.
%
We may also solve for the time-evolution in the Heisenberg picture. Noting that, $\hat{\gamma}_i(t)=\mathcal{S}(t,t_0)\hat{\gamma}_i(t_0)\mathcal{S}^{\dag}(t,t_0)$, we find
\begin{equation}
    \mathcal{S}^{\dag}(t,t_0)\hat{\gamma}_i(t_0)\mathcal{S}(t,t_0)=\hat{\gamma}_i^{H}(t)=V^{T}_{im}(t,t_0)\hat{\gamma}_{m}(t_0)
\end{equation}
where the index ``$H$'' labels operators in the Heisenberg picture.
Now consider the state-space for the ground-state manifold. For a system with two pairs of MZMs, the state vector $|\psi\rangle$, in the even (odd) subspace, is spanned by the basis $\{|00\rangle_{\rm{phys}},|11\rangle_{\rm{phys}}\}$ ($\{|10\rangle_{\rm{phys}},|01\rangle_{\rm{phys}}\}$). 
In the logical basis, it may be described as
\begin{equation}
    |\psi(t)\rangle=\textrm{cos}(\frac{\theta}{2})|0\rangle_{\rm{log}} +\textrm{sin}(\frac{\theta}{2})e^{i\phi}|1\rangle_{\rm{log}} \label{eq:state}
\end{equation}
in the even subspace
where $\theta\in [0,\pi]$ and $\phi\in S_1$. The odd subspace may be described similarly. 
The parity of the subspace may be distinguished by use of the combined parity operator, $\hat{P}=-4\hat{\gamma}_1\hat{\gamma}_2\hat{\gamma}_3\hat{\gamma}_4$, where for $|\psi\rangle$ in the even (odd) subspace, $\hat{P}|\psi\rangle=-1$ $(1)$. \\ 

Utilizing this, we aim to generally describe the state of the system over time by analyzing the time-evolved $\hat{\sigma}_i(t)$ operators. 
We first start with the $t=0$ description of the change of basis matrix, $S(0)=S$:
\begin{equation}
    S=\begin{pmatrix}
                        1 & -i & 0 & 0 \\ 
                        1 & i & 0 & 0\\ 
                        0 & 0 & 1 & -i \\ 
                        0 & 0 & 1 & i
                    \end{pmatrix}.
\end{equation}
With respect to this change of basis matrix, the Pauli matrices $(\hat{\sigma}_x,\hat{\sigma}_y,\hat{\sigma}_z)$ may be represented in the even subspace as $2i(\hat{\gamma}_3\hat{\gamma}_2,\hat{\gamma}_1\hat{\gamma}_3,\hat{\gamma}_2\hat{\gamma}_1)$. 
For the single-qubit Pauli projections, we get that
\begin{align}
    \langle \hat{\sigma}_x\rangle =  \textrm{sin}(\theta) \textrm{cos}(\phi), \quad \langle \hat{\sigma}_y \rangle = \textrm{sin}(\theta) \textrm{sin}(\phi), \quad \langle \hat{\sigma}_z \rangle = \textrm{cos}(\theta)
\end{align}
which will fully describe the many-body state. 
We first define the $n$-MZM product tensor $\hat{\Gamma}^{(n)}_{i_1,i_2,.....,i_{n}}(t)=\prod^{n}_{k=1}\hat{\gamma}_{i_k}(t)$. 
Re-expressing an arbitrary $\hat{\gamma}^H_i(t)=V^{T}_{im}(t)\hat{\gamma}_m$, we find that for a system with $2N$-MZMs, we have
\begin{equation}
    \hat{\Gamma}^{(n)}_{i_1,i_2,.....,i_{n}}(t)=\prod^{n}_{k=1}\sum^{2N}_{j_k=1}V^{T}_{i_kj_k}(t)\hat{\gamma}_{j_k}.
\end{equation}
Focusing on bilinear projections, i.e., $\hat{\Gamma}^{(2)}_{ij}(t)$, and noting Majorana operator anti-commutation relations, i.e., $\{\hat{\gamma}_i(t),\hat{\gamma}_{j}(t)\}=\delta_{ij}$, we find for a system consisting of two pairs of MZMs (N=2):
%
\begin{equation}
    \hat{\Gamma}^{(2)}(t)=\frac{I_{4\times 4}}{2}-\frac{i}{2}V^{T}(t)\begin{pmatrix}
                        0 & -\hat{\sigma}_z & \hat{\sigma}_y  & -\hat{\sigma}_x  \hat{P}  \\ 
                        \hat{\sigma}_z &  0&  -\hat{\sigma}_x  & -\hat{\sigma}_y  \hat{P}  \\ 
                        -\hat{\sigma}_y & \hat{\sigma}_x & 0 & -\hat{\sigma}_z  \hat{P} \\ 
                        \hat{\sigma}_x  \hat{P} & \hat{\sigma}_y  \hat{P} & \hat{\sigma}_z  \hat{P} & 0
                         \end{pmatrix} V(t)\ .
\end{equation}
Thus, to track the time-evolution $|\Psi(t)\rangle$, we obtain the expression
\begin{align}\label{eq:Gammaij}
\begin{split}
         \langle\Gamma^{(2)}_{ij}(t)\rangle=&\frac{1}{2}\delta_{ij}-\frac{i}{2}V^{T}_{ik}(t)\langle \hat{\gamma}_k\hat{\gamma}_l\rangle V_{lj}(t) \\
            =&\frac{1}{2}\delta_{ij}-\frac{i}{2}((V^{T}_{i2}(t)V_{1j}(t)-V^{T}_{i1}(t)V_{2j}(t)-(-1)^p(V^{T}_{i3}(t)V_{4j}(t)-V^{T}_{i4}(t)V_{3j}(t)))\textrm{cos}(\theta)\\[7pt]
            &+\left(V^{T}_{i3}(t)V_{2j}(t)-V^{T}_{i2}(t)V_{3j}(t)-(-1)^p(V^{T}_{i1}(t)V_{4j}(t)-V^{T}_{i4}(t)V_{1j}(t))\right)\textrm{sin}(\theta) \textrm{cos}(\phi)\\[7pt]
            &+\left(V^{T}_{i1}(t)V_{3j}(t)-V^{T}_{i3}(t)V_{1j}(t)-(-1)^p(V^{T}_{i2}(t)V_{4j}(t)-V^{T}_{i4}(t)V_{2j}(t))\right)\textrm{sin}(\theta) \textrm{sin}(\phi)). 
    \end{split}
\end{align}

Here, $p$ labels the parity of the subspace, with $p=0$ $(1)$ indexing the even (odd) subspace. This provides a general method to find the time-evolved wavefunction $|\psi(t)\rangle$ for a single-qubit state in the adiabatic regime.




\section{Braiding Permutations}

Suppose at some time $t'$ in the time-evolution that the basis
expansion coefficients are instantaneously permuted. 
In the quasiparticle basis, the new expansion coefficients $\left(c^P\right)^i_\nu(t)$ may be expressed with respect to the old basis set by 
\begin{equation}
    \left(c^P\right)^i_\mu(t')=P_{\mu \lambda}(t')c^i_\lambda(t')
\end{equation}
where $P(t)$ is a permutation matrix. 
We assume that both \{$\left(c^P\right)^i_\mu(t')$\} and $\{c^i_\mu(t')\}$ may be mapped to the Majorana basis $\{b^i_j(t')\}$ by the change of basis matrices $W^P(t')$ and $W(t')$, respectively. 
Starting from $\left(c^P\right)^i_\mu(t')$, we find that 
\begin{equation}
    \left(c^P\right)^i_\mu(t')=(W^P(t'))_{\mu k}^*b^i_k(t')=\left((W^P(t'))^*W^T(t')\right)_{\mu \nu}c^i_\nu(t')=P_{\mu \nu}(t')c^i_\nu(t').
\end{equation}
giving the required form for the permutation matrix $P(t)$. \\

For the purpose of Majorana braiding, this will occur when one MZM, $\gamma_i$, crosses the worldline of another MZM, $\gamma_j$, at some $t=T_{\textrm{exch}}$, by MZM exchange statistics, if $\gamma_i \to \gamma_j$, $\gamma_j\to -\gamma_i$. This will introduce a discontinuity in the $\{\psi^{\gamma}_i(t)\}$ wavefunction set. 
To preserve continuity over the full evolution in the Majorana wavefunctions, 
we introduce the following $S^P(t')$ matrices about the $t'=T_{\textrm{exch}}$ point for the $\sqrt{\textrm{X}}$ and $\sqrt{\textrm{Z}}$ gate respectively: 

\begin{equation}
    W_{\sqrt{\textrm{X}}}(T_{\textrm{exch}})=                              \begin{pmatrix}
                             1 & 0 & i & 0\\ 
                             1 & 0 & -i & 0\\ 
                             0 & 1 & 0 & -i\\ 
                             0 & 1 & 0 & i
                         \end{pmatrix}, \quad 
    W_{\sqrt{\textrm{Z}}}(T_{\textrm{exch}})=                              \begin{pmatrix}
                             -i & -1 & 0 & 0\\ 
                             -i & 1 & 0 & 0\\ 
                             0 & 0 & 1 & -i\\ 
                             0 & 0 & 1 & i
                         \end{pmatrix}
\end{equation} 
The transfer matrices in each case may then be written as follows:
\begin{align}
    &I^{\sqrt{X}}(T,0)=I(T,T_{\textrm{exch}})W_{\sqrt{\textrm{X}}}^*(T_{\textrm{exch}})W^T(T_{\textrm{exch}})I(T_{\textrm{exch}},0)\\ 
   &I^{\sqrt{Z}}(T,0)=I(T,T_{\textrm{exch}})W_{\sqrt{\textrm{Z}}}^*(T_{\textrm{exch}})W^T(T_{\textrm{exch}})I(T_{\textrm{exch}},0).
\end{align}

\section{Perturbative Approach}

To analytically solve for the $V$-matrix, we begin with the form for $M(t)$ in the $\{|\psi^p_{12}(t)\rangle,|\psi^h_{12}(t)\rangle,|\psi^p_{34}(t)\rangle,|\psi^h_{12}(t)\rangle\}$ basis, 
\begin{align}
    M(t)&=\begin{pmatrix}
           M_{12}(t)& M_{23}(t)\\ 
            M^T_{23}(t) & M_{34}(t)
            \end{pmatrix}=\begin{pmatrix}
           M_{12}(t)& 0\\ 
            0 & M_{34}(t)
            \end{pmatrix}+\begin{pmatrix}
                               0& M_{23}(t)\\ 
                                M^T_{23}(t) & 0
                            \end{pmatrix}\ ,
\end{align}
where 
\begin{eqnarray}
    M_{12}(t) &=& E_{12}(t)\sigma_z +A_{12}(t)\sigma_x\ , \\[5pt]
    M_{34}(t) &=& E_{34}(t)\sigma_z+A_{34}(t)\sigma_x\ ,   \\[5pt]
    M_{23}(t) &=& -(\frac{E_{23}}{2}(t)+A_{13}(t))(\sigma_z+i\sigma_y)\ .
\end{eqnarray}


We take $M^{(1)}(t)=$ to be the diagonal contribution to $M(t)$, with $M^{(2)}(t)$ being the off-diagonal contribution. \\

We consider first the energy-splitting regime.
To do so, we enter the interaction picture, where we make the substitution $c^i_\mu(t)=\mathcal{T}e^{-i\int^t_{t_0}M^{(1)}(t')dt'}(c^i_\mu(t))^I$.   
Solving for $(c^i_\mu(t))^I$, we find that
\begin{align}
    (c^i_\mu(t))^I=\mathcal{T}\textrm{exp}(-i\int^t_0\tilde{\mathcal{T}}e^{i\int^{t'}_{t_0}M^{(1)}(t'')dt''}M^{(2)}(t')\mathcal{T}e^{-i\int^{t'}_{t_0}M^{(1)}(t'')dt''}dt')_{\mu\nu}(c^i_\nu(t_0))^I 
\end{align}
where $\tilde{\mathcal{T}}$ is the anti-time ordering operator.
Suppose now $|M^{(2)}_{ij}(t)| \sim \mathcal{O}(\lambda)$ at all times $t$ for all $i,j$.  We take an expansion of the $c^i_{j}(t)$ term up to $\mathcal{O}(\lambda)$:
\begin{align}
    c^i_\mu(t)&= \mathcal{T}e^{-i\int^{t}_{t_0}M^{(1)}(t')dt'}\left( I_{4 \times 4}-i\int^t_0\Tilde{\mathcal{T}}e^{i\int^{t'}_{t_0}M^{(1)}(t'')dt''}M^{(2)}(t')\mathcal{T}e^{-i\int^{t'}_{t_0}M^{(1)}(t'')dt''}dt'\right)_{\mu\nu}c^i_\nu(t_0)+\mathcal{O}(\lambda^2)\\
    &=\mathcal{T}e^{-i\int^{t}_{t_0}M^{(1)}(t')dt'}\left(I_{4 \times 4}+ \Sigma^{(2)}(t') \right)_{\mu\nu}c^i_{\nu}(t_0)+\mathcal{O}(\lambda^2)
\end{align}
where $\Sigma^{(i)}(t)=-i\int^t_0\Tilde{\mathcal{T}}e^{i\int^{t'}_{t_0}M^{(i)}(t'')dt''}M^{(2)}(t')\mathcal{T}e^{-i\int^{t'}_{t_0}M^{(1)}(t'')dt''}$ with $i\in\{1,2\}$ and $I_{4\times 4}$ is the identity matrix.
The $V$-matrix is then given by 
\begin{align}
    V_{ik}(t,t_0)&=(b^{i}_{j}(t))^TB^T_{jk}(t,t_0)=V^{(1)}_{ik}(t) +V^{(2)}_{ik}(t)+\mathcal{O}(\lambda^2),\label{eq:PerturbativeE-Vmat}
\end{align}
with
\begin{align}
    V^{(1)}_{ik}(t,t_0)&=\left(W^*(t)\mathcal{T}e^{-i\int^{t'}_{t_0}M^{(1)}(t'')dt''}W^T(t_0)\vec{b}(t_0)\right)_{im}B^T_{mk}(t,t_0)\ ,\\ 
    V^{(2)}_{ik}(t,t_0)&=\left(W^*(t)\mathcal{T}e^{-i\int^{t'}_{t_0}M^{(1)}(t'')dt''}\Sigma^{(2)}(t')W^T(t_0)\vec{b}(t_0)\right)^T_{im}B^T_{mk}(t,t_0)\ .
\end{align}

$V^{(1)}_{ik}(t,t_0)$ is the dominant energy splitting contribution to the $V$-matrix, while  $V^{(2)}_{ik}(t)$ is the tunneling contribution. 

We now consider the $\langle \hat{\Gamma}(t) \rangle$ calculation. 
To first order in $\lambda$, the projective calculations onto the Bloch sphere will be given by 

\begin{equation}
\begin{split}
    \langle \hat{\Gamma}^{(2)}(t)\rangle&=\left(V^{(1)}(t,t_0)+V^{(2)}(t,t_0)+\mathcal{O}(\lambda^2)\right)^{-1}\langle \hat{\Gamma}^{(2)}(t_0)\rangle\left(V^{(1)}(t,t_0)+V^{(2)}(t,t_0)+\mathcal{O}(\lambda^2)\right)\\
&\overset{\lambda \to 0}{\to} (V^{(1)}(t,t_0))^{-1}\langle\hat{\Gamma}^{(2)}(t_0)\rangle V^{(1)}(t,t_0).
\end{split}
\end{equation}
 This gives the leading order contribution to the single-qubit correlation tensor in the energy-splitting regime.  
 
 If we take $|M^{(2)}_{\mu\nu}(t)|\sim \lambda$ the leading order contribution in the tunneling regime may also be found by interchanging the labels $(1)$ and $(2)$, i.e., 
\begin{equation}
\begin{split}
    \langle \hat{\Gamma}^{(2)}(t)\rangle 
\overset{\lambda \to 0}{\to} (V^{(2)}(t,t_0))^{-1}\langle\hat{\Gamma}^{(2)}(t_0)\rangle V^{(2)}(t,t_0)
\end{split}
\end{equation}
which is the leading order term in the tunneling regime. 
The $V^{(i)}$ matrices in this case are given by
\begin{align}
    V^{(2)}_{ik}(t,t_0)&=\left(W^*(t)\mathcal{T}e^{-i\int^{t'}_{t_0}M^{(2)}(t'')dt''}W^T(t_0)\vec{b}(t_0)\right)^T_{im}B^T_{mk}(t,t_0)\ , \\ 
    V^{(1)}_{ik}(t,t_0)&=\left(W^*(t)\mathcal{T}e^{-i\int^{t'}_{t_0}M^{(2)}(t'')dt''}\Sigma^{(1)}(t')W^T(t_0)\vec{b}(t_0)\right)^T_{im}B^T_{mk}(t,t_0).
\end{align}

\subsection{Scaling}
In the limiting regimes, the single-qubit projections may be found by considering the quantities $\int^t_0E_\mu(t')dt'$ and $\int^t_0A_{\mu \nu}(t')dt'$. 

Suppose for two identical braiding protocols, we have two total braid times $T_1$, $T_2$, where $0 \leq t_{1} \leq T_{1}$, $0 \leq t_{2} \leq T_{2}$ and $T_2=kT_1$ with $k\in\mathbb{R}$. 
We may also identify $t_1$ with $t_2$ by $t_2=kt_1$.

Consider a system $\tilde{H}_{\textrm{BdG}}(\tau)$ with  total braid time $T=1$.  
Time in this system is given by the dimensionless quantity $\tau=t_1/T_1=t_2/T_2$, whereby $0 \leq \tau \leq 1$. 
As such, we may arbitrarily map to any time $t_i$ by $t_i=T_i\tau$. 
For $\tilde{H}_{\textrm{BdG}}(\tau)$ the BdG Hamiltonian for a system with total braid time 1, and $H^i_{\textrm{BdG}}(t_i)$ the BdG Hamiltonian for a system with total braid time $T_i$, there exists a mapping
between their respective eigenvector sets $\{|\tilde{\psi}_\mu(\tau)\rangle\}=\{|\psi_\mu(t_i)\rangle\}$ at time $t_i=T_i\tau$. \\

With this connection between BdG eigenstates, we may define the dimensionless non-Abelian Berry-matrix, $\tilde{A}_{\mu\nu}(\tau)$, as
\begin{equation}
    \tilde{A}_{\mu\nu}(\tau)=i\langle \tilde{\psi}_{\mu}(\tau)|\dot{\tilde{\psi}}_{\nu}(\tau)\rangle.
\end{equation}

We recognize that for $t=T\tau$, we may associate $A_{\mu\nu}(t)$ with $\tilde{A}_{\mu\nu}(\tau)$ by 
\begin{equation}
    A_{\mu\nu}(t)=i\langle \psi_\mu(T\tau)|\overset{\rightarrow}{\partial}_{T\tau}|\psi_\nu(T\tau)\rangle=\frac{1}{T}\langle \tilde{\psi}_{\mu}(\tau)|\overset{\rightarrow}{{\partial}_{\tau}}|\tilde{\psi}_{\nu}(\tau)\rangle=\frac{i}{T}\tilde{A}_{\mu\nu}(\tau).
\end{equation}

First consider the integral over Berry-matrix elements 
\begin{equation}
    \int^t_0A_{\mu\nu}(t')dt'=\int^{t}_0\frac{1}{T}\tilde{A}_{\mu\nu}(\tau)d(T\tau)=\int^{1}_0\tilde{A}_{\mu\nu}(\tau)d\tau\ , 
\end{equation}
i.e., $\int^t_0A_{\mu\nu}(t')dt'\to \beta$ in the adiabatic limit, where $\beta$ is a constant. 
Next, consider the energy integral 
\begin{equation}
    \int^t_0E_\mu(t')dt'=\int^{t}_0\tilde{E}_\mu(\tau)d(T\tau)=T\int^{1}_0\tilde{E}_\mu(\tau)d(\tau) =T\bar{E}_\mu.
\end{equation}
As such, the energy contribution scales with the braiding time $T$.

\subsection{V-matrices}
In the energy-splitting and tunneling regimes, the
single-qubit V-matrices are listed below, with associated average energy $\bar{E}_{ij}$ and integrated non-Abelian Berry matrix $\beta_{ij}$ contributions. 

\vspace{10pt}
\underline{Energy-Splitting Regime:}\\
\begin{align}
    (V^{T})^{\textrm{split}}_{\sqrt{\textrm{X}}}&=\begin{pmatrix}
                            \textrm{cos}(\bar{E}_{12}T)\textrm{cos}(\bar{E}_{13}T) & \textrm{cos}(\bar{E}_{13}T)\textrm{sin}(\bar{E}_{12}T) & -\textrm{cos}(\bar{E}_{34}T)\textrm{sin}(\bar{E}_{13}T) & -\textrm{sin}(\bar{E}_{13}T)\textrm{sin}(\bar{E}_{34}T)\\[5pt] 
                            -\textrm{cos}(\bar{E}_{12}T)\textrm{sin}(\bar{E}_{13}T) & -\textrm{sin}(\bar{E}_{12}T)\textrm{sin}(\bar{E}_{13}T) & -\textrm{cos}(\bar{E}_{13}T)\textrm{cos}(\bar{E}_{34}T) & -\textrm{cos}(\bar{E}_{13}T)\textrm{sin}(\bar{E}_{34}T)\\[5pt] 
                            -\textrm{cos}(\bar{E}_{24}T)\textrm{sin}(\bar{E}_{12}T) & \textrm{cos}(\bar{E}_{12}T)\textrm{cos}(\bar{E}_{24}T) & -\textrm{sin}(\bar{E}_{24}T)\textrm{sin}(\bar{E}_{34}T) & \textrm{cos}(\bar{E}_{34}T)\textrm{sin}(\bar{E}_{24}T)\\[5pt] 
                            \textrm{sin}(\bar{E}_{12}T)\textrm{sin}(\bar{E}_{24}T) & -\textrm{cos}(\bar{E}_{12}T)\textrm{sin}(\bar{E}_{24}T) & -\textrm{cos}(\bar{E}_{24}T)\textrm{sin}(\bar{E}_{34}T) & \textrm{cos}(\bar{E}_{24}T)\textrm{cos}(\bar{E}_{34}T)
                            \end{pmatrix}
\end{align}

\begin{align}
    (V^{T})^{\textrm{split}}_{\sqrt{\textrm{Z}}}&=\begin{pmatrix}
                                    \textrm{sin}(\bar{E}_{12}T) & -\textrm{cos}(\bar{E}_{12}T) & 0 & 0\\[5pt]  
                                    \textrm{cos}(\bar{E}_{12}T) & \textrm{sin}(\bar{E}_{12}T) & 0 & 0\\[5pt] 
                                    0 & 0& \textrm{cos}(\bar{E}_{34}T) & \textrm{sin}(\bar{E}_{34}T)\\[5pt] 
                                    0 & 0& -\textrm{sin}(\bar{E}_{34}T) & \textrm{cos}(\bar{E}_{34}T)
                                \end{pmatrix}\\[15pt] 
    (V^{T})^{\textrm{split}}_{R_{\textrm{Z}}(\phi)}&=\begin{pmatrix}
                                    \textrm{cos}(\bar{E}_{12}T) & \textrm{sin}(\bar{E}_{12}T) & 0 & 0\\[5pt]  
                                    -\textrm{sin}(\bar{E}_{12}T) & \textrm{cos}(\bar{E}_{12}T) & 0 & 0\\[5pt] 
                                    0 & 0& \textrm{cos}(\bar{E}_{34}T) & \textrm{sin}(\bar{E}_{34}T)\\[5pt] 
                                    0 & 0& -\textrm{sin}(\bar{E}_{34}T) & \textrm{cos}(\bar{E}_{34}T)
                                \end{pmatrix}
\end{align}

\vspace{10pt}
\underline{Tunneling Regime:}
\begin{align}
    (V^{T})^{\textrm{tun}}_{\sqrt{\textrm{X}}}&=\begin{pmatrix}
                                    1 & 0 & 0 & 0\\[5pt] 
                                    0 & \textrm{sin}(\bar{E}_{23}T+2\beta) & -\textrm{cos}(\bar{E}_{23}T+2\beta) & 0\\[5pt]
                                    0 & \textrm{cos}(\bar{E}_{23}T+2\beta)& \textrm{sin}(\bar{E}_{23}T+2\beta) & 0\\[5pt]
                                    0 & 0& 0 & 1
                                \end{pmatrix}\\[15pt]
    (V^{T})^{\textrm{tun}}_{R_{\textrm{X}}(\phi)}&=\begin{pmatrix}
                                    1 & 0 & 0 & 0\\[5pt] 
                                    0 & \textrm{cos}(\bar{E}_{23}T+2\beta) & \textrm{sin}(\bar{E}_{23}T+2\beta) & 0\\[5pt]
                                    0 & -\textrm{sin}(\bar{E}_{23}T+2\beta)& \textrm{cos}(\bar{E}_{23}T+2\beta) & 0\\[5pt] 
                                    0 & 0 & 0 & 1
                                \end{pmatrix}
\end{align}

We list in Table\,\ref{tab:VMats} the required $V$-matrices to reproduce each corresponding one-qubit gate.  
State evolution as a function of braid time $T$ may be analytically tracked by solving for the required time-dependent Pauli-operators $\left(\hat{\sigma}_x(T),\hat{\sigma}_y(T),\hat{\sigma}_{z}(T))=2i(\hat{\Gamma}^{(2)}_{32}(T),\hat{\Gamma}^{(2)}_{13}(T),\hat{\Gamma}^{(2)}_{21}(T)\right)$. 

\begin{table}[h!]
    \centering
    \begin{tabular}{c|c}
    Gate & Inverse V-Matrix Composition $V^{T}(t)$ \\[3pt] 
    \hline 
    $\sqrt{\textrm{X}}$ & $V^{T}_{\sqrt{\textrm{X}}}$ \\[3pt] 
    \hline
    $\textrm{X}$ & $V^{T}_{\sqrt{\textrm{X}}}V^{T}_{\sqrt{\textrm{X}}}$ \\[3pt]
    \hline 
    $\sqrt{\textrm{Z}}$ & $V^{T}_{\sqrt{\textrm{Z}}}$ \\[3pt] 
    \hline
    $\textrm{Z}$ & $V^{T}_{\sqrt{\textrm{Z}}}V^{T}_{\sqrt{\textrm{Z}}}$ \\[3pt]
    \hline
    $\textrm{H}$ & $V^{T}_{\sqrt{\textrm{Z}}}V^{T}_{\sqrt{\textrm{X}}}V^{T}_{\sqrt{\textrm{Z}}}$ \\[3pt]
    \hline
    $R_{\textrm{X}}(\phi)$ & $(V^{T})^{\textrm{tun}}_{R_{\textrm{X}}(\phi)}$\\[3pt]
    \hline
    $R_{\textrm{Y}}(\phi)$ & $V^{E_{ij} \to 0}_{\sqrt{\textrm{X}}}(V^{T})^{\textrm{split}}_{R_{\textrm{Z}}(\phi)}(V^{T})^{E_{ij} \to 0}_{\sqrt{\textrm{X}}}$\\[3pt]
    \hline
    $R_{\textrm{Z}}(\phi)$ & $(V^{T})^{\textrm{split}}_{R_{\textrm{Z}}(\phi)}$
    \end{tabular}
    \caption{Single-qubit Clifford gates with the associated $V$-matrices needed to reproduce the gate. Additionally, arbitrary rotation operators about the $(x,y,z)$ axes on the Bloch sphere $R_{\textrm{X}}(\phi)$, $R_{\textrm{Y}}(\phi)$, $R_{\textrm{Z}}(\phi)$ are also provided.}
    \label{tab:VMats}
\end{table}

\subsection{Example 1: X-Gate as a Function of MZM-Separation}

\begin{figure}[t!]
    \centering
    \includegraphics{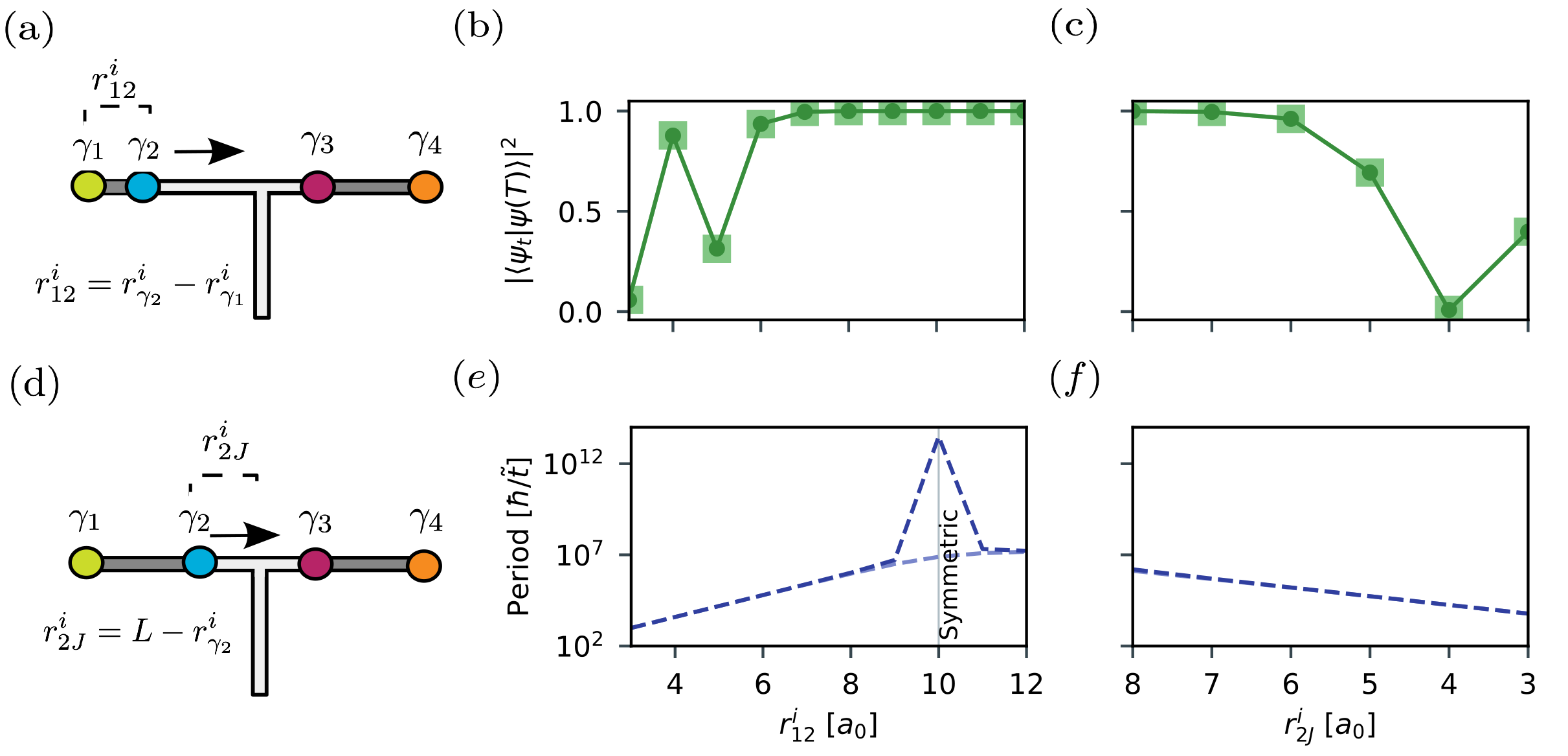}
    \caption{
    (a) Graphic of a T-junction set-up with initial spatial separation of $\gamma_1$ and $\gamma_2$ given by $r^i_{12}=r^i_{\gamma_2}-r^i_{\gamma_1}$. (b) Fidelities, $|\langle 1|0(t)\rangle_{\rm log}|^2$, for fixed braid time $2T=3600 \hbar/t$, are plotted as a function of $r^i_{12}$. (c) The period of oscillation. $\tau$, is plotted as a function of distance $r^i_{12}$. Both even and odd regimes considered with $r^i_{34}=6a_0$ kept fixed. 
    (d) Same as (a) with $r^i_{2J}=L-r^i_{\gamma_2}$ being the separation between $\gamma_2$ and the junction. 
    %
    (e, f): Same as (b, c) but replacing $r^i_{12}$ with $r^i_{2J}=L-r^i_{\gamma_2}$ for the tunneling regime.
    $a_0$ is the lattice spacing.}
    \label{fig:Fig2sup}
\end{figure}
Additional to the examples given in the main text, we also include the example in Fig.\,\ref{fig:Fig2sup} of Pauli X-gate success probability, otherwise known as the fidelity, as a function of Majorana separation. 

Here, we fix the position of $\gamma_1$, $\gamma_3$ and $\gamma_4$ and move the initial position of $\gamma_2$ along the first junction of the leg. 
As shown in Fig.\,\ref{fig:Fig2sup}\,a, initially $\gamma_2$ is closest to $\gamma_1$ with the relevant length scale in this case being $r^i_{12}=r^i_{\gamma_2}-r^i_{\gamma_1}$, where $r^i_{\gamma_i}$ is the initial position of $\gamma_i$. 
Starting from $|0\rangle_{\rm log}$, we measure the fidelity $|\langle 1|0(t)\rangle_{\rm log}|^2$ as given in Fig.\,\ref{fig:Fig2sup}\,b which reveals a stable X-gate with $|\langle 1|0(T)\rangle_{\rm log}|^2 \to 1$ in the regime with $r^i_{12}\geq 8 a_0$. 
Further, the associated period of oscillation between a success ($|0\rangle_{\rm log}\to |1\rangle_{\rm log}$) and a fail ($|0\rangle_{\rm log} \to |0\rangle_{\rm log}$) is shown in Fig.\,\ref{fig:Fig2sup}\,c. 
Initially, as the dominant overlap is through the topological region, the system remains in the energy splitting regime, with  $\langle\hat{\sigma}_z\rangle=\textrm{cos}\left((\bar{E}_{12}+\bar{E}_{13}+(-1)^p(\bar{E}_{34}+\bar{E}_{24}))T)\right)$, where $p=0$ $(1)$ for the even (odd) subspace.
We reveal a potential advantage with conducting the X-gate in the odd subspace, with the period of oscillation longer in this regime, leading to larger time widths of potential success. 
This is emphasized when the separation positions of all MZMs $r^i_{12}=r^i_{34}=10 a_0$ is entirely symmetric, leading to $\bar{E}_{12}+\bar{E}_{13}-\left(\bar{E}_{34}+\bar{E}_{24}\right)\to 0$ in the odd regime, while the even regime maintains a large but finite oscillation period.

As the initial $\gamma_2$ position approaches the junction, the dominant overlap throughout the process will occur between $\gamma_2$ and $\gamma_3$, thus entering the tunneling regime. Similarly, as given in Fig.\,\ref{fig:Fig2sup}\,d,  we plot transition probabilities from $r^i_{2J}=8 a_0$, showing a successful braid until $r^i_{2J}=6 a_0$ whereby the success probability begins to perturb away from 1 (Fig.\,\ref{fig:Fig2sup}\,e). 
We can measure the period as well in this regime, as shown in Fig.\,\ref{fig:Fig2sup}\,f where the angular oscillation frequency $\omega = \bar{E}_{23}$ as $\bar{E}_{14}\to 0$ for large $\gamma_1-\gamma_4$ separation in both the odd and even subspaces. 
As such, the angular oscillation frequency in both regimes is the same.

Finally, when $r_{12}$ ($r_{2J}$) increases, $E_{12}(t)\sim e^{-r_{12}(t)/\zeta(t)}$ ($E_{23}(t)\sim e^{-r_{23}(t)/\zeta(t)}$)\,\cite{cheng2009} changes exponentially as a function of Majorana separation.
This is validated in both Fig.\,\ref{fig:Fig2sup}\,c and f which show linear growth and decay, respectively, as a function of $r^i_{12}$ and $r^i_{2J}$ with respect to the logarithmic period scaling.

\subsection{Example 2: Additional Gates on a T-Junction}

\begin{figure}[t!]
    \centering
    \includegraphics{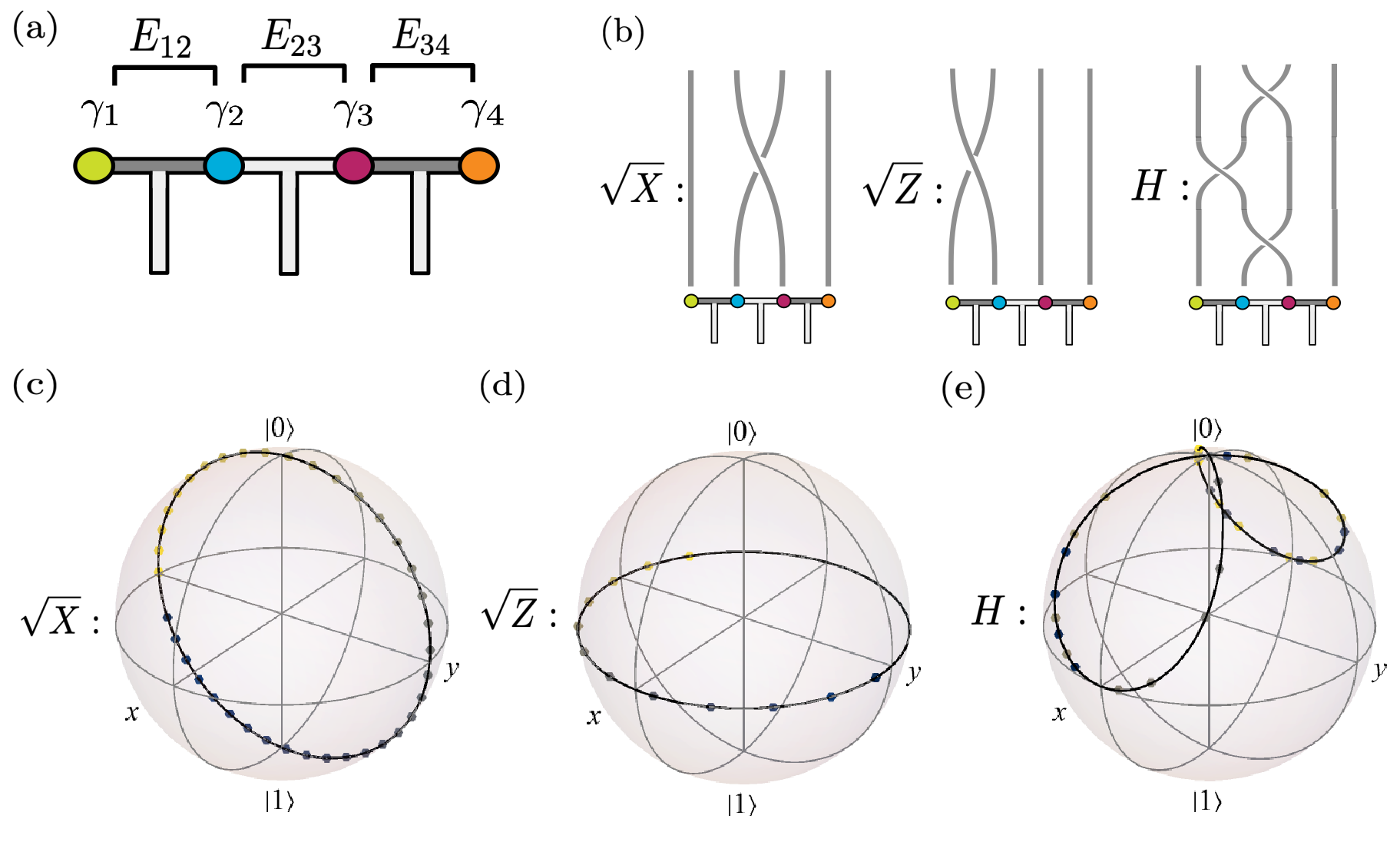}
    \caption{State oscillations of additional gates: (a) Graphic of a tri-T junction. (b) Graphic of the requisite world-lines for a $\sqrt{X}$, $\sqrt{Z}$, $H$ gate. (c), (d), (e) The final state outcomes for a $\sqrt{X}$, $\sqrt{Z}$ and $H$ braid respectively, with analytical final state trajectories (solid line) plotted against many-body simulations (dots).  
    The parameter set for each case are ($\mu_{\textrm{topo}}$, $\mu_{\textrm{triv}}$, $r_{12}$, $r_{34}$, $L$)=$(0.55,5.5,14a_0,15a_0,8a_0)$, ($\mu_{\textrm{topo}}$, $\mu_{\textrm{triv}}$, $r_{12}$, $r_{34}$, $L$)=$(0.7,7.6,14a_0,15a_0,7a_0)$, ($\mu_{\textrm{topo}}$, $\mu_{\textrm{triv}}$, $r_{12}$, $r_{34}$, $L$)=$(0.15,7.2,15a_0,13a_0,6a_0)$ respectively.}
    \label{fig:Fig4sup}
\end{figure}

We emphasize the results given in Tab. \ref{tab:VMats} by demonstrating oscillations on a tri-T junction, as given in Fig. \ref{fig:Fig4sup} (a).
Such a model has the distinct advantage that we may disentangle the hybridization regimes by the addition of auxiliary legs between each MZM. 
This means, assuming $E_{ij}\sim 0$ between all MZMs initially, that we may focus on hybridization between a single pair of MZMs $\gamma_i$, $\gamma_j$ involved in a braid, during which the MZMs are closest to each other. 
Here, we analyze a $\sqrt{X}$, $\sqrt{Z}$ and $H$-gate, with the world lines corresponding to each gate given in \ref{fig:Fig4sup} (b). 
The final state outcomes and state trajectories are given for each gate respectively in Fig. \ref{fig:Fig4sup} (c)-(e). \\

We first focus on the $\sqrt{X}$-gate and $\sqrt{\rm{Z}}$-gate.
As the dominant energy coupling is $E_{23}$ ($E_{12}$), we may solve each case in the tunneling and energy-splitting regime respectively. 
As $e^{i\theta}|0\rangle \sim |0\rangle_{\rm{log}}$, for the $\sqrt{\rm{Z}}$-gate, we initialize $|\psi\rangle=\frac{1}{\sqrt{2}}\left(|0\rangle_{\rm{log}}+|1\rangle_{\rm{log}}\right)$ to see the phase evolution after the braiding operation. 
We set $|\psi\rangle=|0\rangle_{\rm{log}}$ as usual for the $\sqrt{X}$-gate. 
In each case, we reveal a steady oscillation about the $x$ and $z$-axis respectively, with period of oscillation $T=\frac{2\pi}{\bar{E}_{23}}$ and $\frac{2\pi}{\bar{E}_{12}}$. 
The $\sqrt{\rm{Z}}$-gate case has been analyzed previously in \cite{cheng2011}, in the case of a constant energy splitting, however here, we provide the result in the context of a full dynamic calculation.\\

However, we emphasize the versatility of the method in Fig. \ref{fig:Fig4sup} (e). 
As indicated in Tab. \ref{tab:VMats}, we may gain the corresponding $V$-matrix for a Hadamard gate by the composition of $V$-matrices $V^T_H=V^{T}_{\sqrt{\textrm{Z}}}V^{T}_{\sqrt{\textrm{X}}}V^{T}_{\sqrt{\textrm{Z}}}$. 
As such, with the building blocks established in Fig. \ref{fig:Fig4sup} (c) and (d), we may compose these results to analyze the final state trajectory of a Hadamard gate. 
We see that in the case given here of large $E_{23}$ vs. $E_{12}$, the state, initialized at $|0\rangle$, moves about the top sector of the Bloch-sphere. 
This establishes the strength of the method, as we may clearly analyze error accumulated by more complicated braiding operations by composition. 
This is additionally also done in the main text to analyze the $\rm{X}$-gate ($V^T_X=\left[V^T_{\sqrt{X}}\right]^2$) and the $R_Y$-gate ($V^T_{R_{\rm Y}(\phi)}=V^{E_{ij} \to 0}_{\sqrt{\textrm{X}}}(V^{T})^{\textrm{split}}_{R_{\textrm{Z}}(\phi)}(V^{T})^{E_{ij} \to 0}_{\sqrt{\rm X}}$). 
Further, this is not restricted to single-qubit results, with many-qubit braids subject to the same analysis, offering a powerful method to analyzing error in many-MZM systems.


\subsection{Example 3: X-gate Oscillations of a Spinful T-Junction Model}

\begin{figure}[t!]
    \centering
    \includegraphics{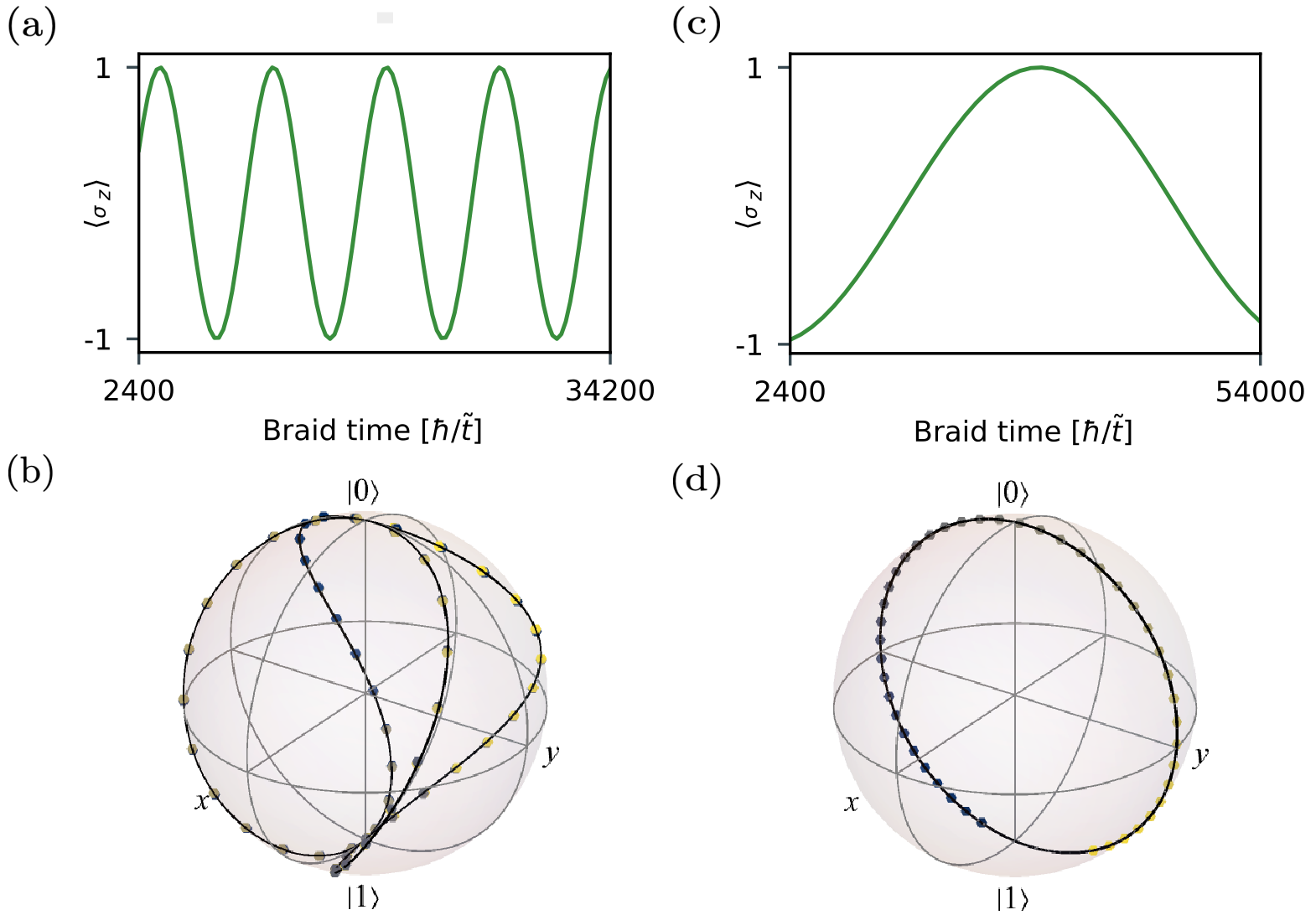}
    \caption{X-gate oscillations due to energy-splitting (first column), tunneling (second column) in the spinful case:
    (a), (c) Pauli-Z projection, $\langle \hat{\sigma}_z(T)\rangle$, as a function of braiding time $T$. (b), (d) Plotted on a Bloch-sphere, the final state outcomes, $|\Psi(T)\rangle$, as a function of total braid time for the X Gate, with analytical final state trajectories (solid line) plotted against many-body simulations (dots).  
    The parameter set for each case are ($\mu_{\textrm{topo}}$, $\mu_{\textrm{triv}}$, $V_z$, $r_{12}$, $r_{34}$, $L$)=$(1.65,7.35,-1.35,4a_0,4a_0,20a_0)$, ($\mu_{\textrm{topo}}$, $\mu_{\textrm{triv}}$, $V_z$, $r_{12}$, $r_{34}$, $L$)=$(1.65,7.35,-1.64,7a_0,17a_0,24a_0)$ respectively, and initial state $|\psi\rangle=|0\rangle_{\rm log}$.} 
    \label{fig:Fig3sup}
\end{figure}

We emphasize the generality of the method by investigating braiding outcomes on the a spinful $p$-wave T-junction.
Similar to the spinless case, the T-junction may be constructed by attaching three one-dimensional legs, as in Fig. \ref{fig:Fig1sup}. 
The Hamiltonian for the one-dimensional wire is given by
\begin{equation}
H = -\mu \sum_{a,\sigma} c_{a,\sigma}^\dagger c_{a,\sigma}^\pd
- \tilde{t} \sum_{a,\sigma} \left(
    c_{a,\sigma}^\dagger c_{a+1,\sigma}^\pd + \hc
\right)+ \sum_{a,\sigma,\sigma'} (V_z c_{a\sigma}^        \dagger (\sigma_z)_{\sigma\sigma'} c_{a,\sigma'}^\pd
        +\Delta_p c_{a,\sigma}^\dagger (\sigma_z)_{\sigma\sigma'} c_{a+1,\sigma'}^\dagger + \hc
        ). 
\end{equation}
The additional spin degree of freedom effectively introduces two spinless Kitaev chains, each  in a different spin channel. 
The condition for each to host an MZM is unchanged, with $|\mu|<2t$ for $|\Delta|=t$, with 2 MZMs formed in each spin channel. 
The out of plane Zeeman field $V_z$ breaks the Kramers spin degeneracy, inducing an effective chemical potential in each channel of $\mu_{\rm eff}=\mu \pm V_z$.
For $|\Delta|=|t|=1$ and $V_z<0$, each spin channel now enters a topological phase for $\mu\pm V_z<2$ \cite{Setiawan2017}.
We choose $\mu>0$ and $V_z<0$ such that only a single pair of MZMs form on the boundary of the topological region.\\

We first track the X-gate braiding outcomes in the energy splitting regime as the main text.  
Initializing again with logical $|0\rangle$, Fig \ref{fig:Fig3sup}. (a), (c). recovers the same oscillatory behavior in $\langle \sigma_z(t)\rangle$ for both hybridization channels, a proxy for the fidelity $|\langle 1|0(t)\rangle|^2$. 
This correspondence also holds when analyzing the phase, which we can see in Fig. \ref{fig:Fig3sup}. (b), (d). 
Much like in the spinless case, the final state outcomes match the analytical relations given in Eq. (5)-(10) in the main text. 
These result emphasize the point made in the main text, that the methods used here are independent on the degrees of freedom of the model. 
While key results have been confirmed via simulations on the spinless, $p$-wave T-Junction, the same methodology may be used to analyze and benchmark braiding outcomes on more complicated and experimentally feasible structures.


\subsection{Example 4: Pauli-Qubit Error}
\begin{figure}[t!]
    \centering
    \includegraphics[width=0.99\textwidth]{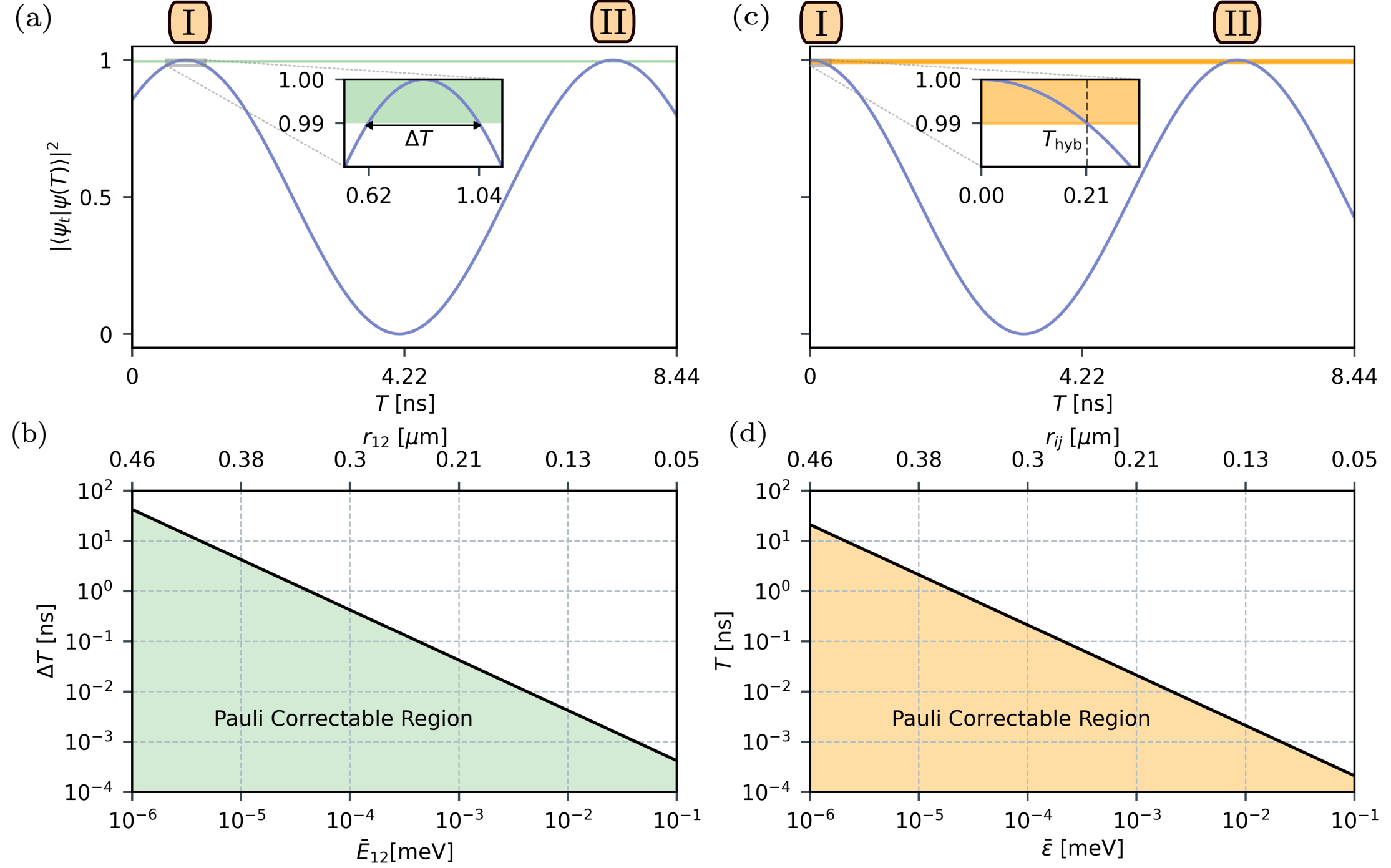}
    \caption{
    Pauli-Qubit Error: {(a) Fidelities, $|\langle \psi_t|\psi(T)\rangle|^2$, are plotted with respect to a chosen target state, $|\psi_t\rangle$, for an R$_z$($\frac{\pi}{4}$) magic gate as a function of $T$. 
   Here, the green region gives the allowed uncertainty in hybridization time, $\Delta T$, where the result remains Pauli correctable.
    (c) Same as a but for an X gate. The insets in a and c show the fidelities about the first Pauli correctable regime, with average energies $\bar{E}_{12}=\bar{\epsilon}=0.1\mu$eV used for a and c.
    (b) Plot of the hybridization times $T$ where the result remains Pauli correctable, against $\bar{E}_{12}$ for the R$_z$($\frac{\pi}{4}$) gate from a about the first Pauli-correctable region. 
    (d) Similar as b but for the X gate: braid time $T$ against the average sum hybridization energies, $\bar{\varepsilon}=\bar{E}_1+\bar{E}_2$, which give the angular frequency of the oscillation. The orange region in (c) and (d) corresponds to the allowed braid times $T<T_{\rm{hyb}}$ where the result remains Pauli correctable.
    In both b and d, each energy is associated with an MZM separation $r_{ij}$, calculated on a superconductor-semiconductor heterostructure, with parameters taken from\,\cite{Pan2024}.}  }
    \label{fig:Fig5sup}
\end{figure}
Finally, we stress how this method may be used to place sensitive bounds on the braid times for which the qubit gates remain Pauli-correctable, required for a fault-tolerant quantum computing platform. 
For a given time-evolution $\mathcal{S}(T,0)$, designed to map an initial state $|\psi\rangle$ to a target state $|\psi_t\rangle$ we set the error correction threshold as $|\langle \psi_t|\psi(T)\rangle|^2>1-\epsilon$, with $\epsilon$ setting the allowed error in the fidelity whereby fault-tolerance is maintained.

For the initial state $|\psi\rangle=\frac{1}{\sqrt{2}}\left(|0\rangle_{\rm{log}} +|1\rangle_{\rm{log}} \right)$ we consider an R$_z$($\frac{\pi}{4}$) gate and show the fidelity $|\langle \psi_t|\psi(T)\rangle|^2$ plotted against hybridization time $T$  
with $|\psi_t\rangle=\frac{1}{\sqrt{2}}\left(|0\rangle_{\rm{log}} +e^{-i\frac{\pi}{4}}|1\rangle_{\rm{log}} \right)$ in Fig.\,\ref{fig:Fig5sup}\,(a).
We set $\bar{E}_{12}= 0.1$$\mu$eV rather than using arbitrary units,  to draw a clear relation between possible experimental energy scales and the allowed hybridization times $T$.
For these choices of initial and target state, we satisfy the correctable error threshold condition when $\left(\rm{cos}(\bar{E}_{12}T)+\rm{sin}(\bar{E}_{12}T)\right)>\sqrt{2}\left(1-2\epsilon \right)$. 
We set $\epsilon=0.01$, well within Pauli-qubit error bounds considered in previous literature \cite{Raussendorf2007,Stace2009}.
This reveals two key upsides:  
Firstly, since $\bar{E}_{12}$ is a dynamic measure of the hybridization throughout the entire routine (as mentioned in the main text), it indeed acts as a sensitive measure for the allowed hybridization times whereby the result remains within fault-tolerance. 
Further, as a consequence of the oscillatory behavior in fidelity, if this time is overshot, the result may be recovered $\delta T=\frac{2\pi}{\bar{E}_{12}}$ later in time.
This is demonstrated in Fig.\,\ref{fig:Fig5sup}\,(a), labelling the first two correctable regimes $\left(\rm{I}\right)$ and $\left(\rm{II}\right)$, respectively.
Secondly, the inset centers on the first Pauli-correctable region, revealing a span of time $\Delta T$ for which the result remains Pauli-correctable. 
As the fidelity is periodic, this time span will be the same for all subsequent Pauli-correctable regions.

We then look to analyze how $\Delta T$ changes as a function of $\bar{E}_{12}$, as done in Fig.\,\ref{fig:Fig5sup}\,(b).
While this reveals relatively minor values in the allowed uncertainty in hybridization time at small average energies, with $\Delta T \sim \mathcal{O}(1\rm{ns})$ for $\bar{E}_{12} \sim \mathcal{O}(0.01\mu \rm{eV})$, we clearly see the advantage that $\Delta T$ scales inversely with $\bar{E}_{12}$, increasing the allowed hybridization time for smaller average energy splittings.
We test this on an experimentally viable platform, the  semiconductor-superconductor (SM-SC) heterostructure, on a parameter set close to experimental values for an InAs-Al nanowire heterostructure, which will host a Majorana bound state \cite{microsoft23,Pan2024}.
Here, we see while the energies are significantly smaller than the gap size $\left(\approx 0.1\rm{meV}\right)$, we can associate each energy splitting with an MZM spatial separation $r_{ij}$ smaller than the length of the nanowire ($\sim 3\mu$m). 
For example, consider an R$_z(\frac{\pi}{4})$ gate protocol where there is a machine uncertainty in the hybridization time $\Delta T_{\rm{exp}}=1\rm{ns}$. 
While this threshold is hit for an average energy $\bar{E}_{12}\approx 0.042 \mu$eV, as can be seen in Fig.\,\ref{fig:Fig5sup}\,(b), $\Delta T_{\rm{exp}}$ will lie below the Pauli correctable region as long as $r_{12}>0.328\mu$m throughout the entire process. 
Thus, while not topologically protected, we may clearly engineer the process which satisfies fault-tolerance on an experimentally viable platform.

This analysis may similarly be used to place bounds on the braid time of a topologically protected gate. 
For initial $|\psi\rangle=|0\rangle_{\rm{log}}$ and $|\psi_t\rangle=|1\rangle_{\rm{log}}$, we again consider the X-gate, setting $\bar{\varepsilon}=\bar{E}_1+\bar{E}_2$, with $\bar{E}_1=\bar{E}_{12}+\bar{E}_{34}$ and $\bar{E}_2=\bar{E}_{13}+\bar{E}_{24}$ as defined for Eq.\,(5-7) in the main text.
Much like in Fig. \ref{fig:Fig5sup} (a), the fidelity as a function of time is plotted against the Pauli correctable region in Fig.\,\ref{fig:Fig5sup}\,(c).
We emphasize the similarity between oscillation periods in (a) and (c) is simply because we set $\bar{E}_{12}=\bar{\varepsilon}=0.1\mu$eV in this case.
We see that, similarly to the R$_z$ gate, we gain a bound for the fidelity, where here, for $\rm{cos}^2\left(\frac{\bar{\varepsilon}T}{2}\right)>0.99$, the computation remains correctable.
Setting the maximal allowed braiding time $T_{\rm{hyb}}$ to be where $\rm{cos}^2\left(\frac{\bar{\varepsilon}T_{\rm{hyb}}}{2}\right)=0.99$, this sets a boundary for the allowed braid time for a suitably topologically protected gate. 
This maximal hybridization time is plotted in Fig. \ref{fig:Fig5sup} (d), with average energy splitting plotted against MZM separation on the SM-SC heterostructure platform.
We see that, similarly to the R$_z$ gate, the time-scale $T_{\rm{hyb}}$ may be optimized by controlling the minimal separation between the MZMs.
For example, we see that for  $r_{ij} >0.38\mu m$, possible with respect to the 3$\mu$m nanowire considered in \cite{microsoft23}, $T_{\rm{hyb}}>1\rm{ns}$, providing a minimal MZM separation where the allowed braiding time remains within a specified machine precision.

This presents a striking similarity between the functioning of the topologically protected X-gate to the timing based R$_z$ gate. 
Here, both $T_{\rm{hyb}}$ and $\Delta T$ may be increased (decreased) by increasing (reducing) the minimum MZM separation over the process. 
This emphasizes that, not only is the topological advantage enhanced as we increase the size of the platform, but also the feasibility of implementing non-topologically protected gates, with the uncertainty allowed for the magic gate, $\Delta T$, increases as a function of the minimal MZM separation.
While demonstrated for these examples, we may similarly analyze bounds for any such qubit-gate within this formalism.





\section{Time-Dependent Pfaffian (TDP) Method}

For full many-body simulations, we make use of the method introduced in \cite{mascot2023}. 
Suppose we have a time-dependent Fock-state $|\vec{n}_d(t)\rangle=\mathcal{S}(t)|\vec{n}_d\rangle$. Defining $|\vec{n}_d\rangle=\prod_k \left( d_k^\dag \right)^{n_k} \ket{0_d}$, where $|0\rangle_d$ corresponds to the superconducting quasiparticle vacuum and $n_k\in\{0,1\}$ defining the occupancy of the single-particle state, the time-evolved Fock-state $\ket{\vec{n}_d}$ is given by
\begin{equation}
\begin{split}
    \ket{\vec{n}_d} = \prod_k \left( d_k^\dag \right)^{n_k} \ket{0_d}&=\prod_k \left( d_k^\dag \right)^{n_k}\prod_{k \in P} \frac{1}{\sqrt{\mathcal{N}}} \prod_{k \in P} \bar{d}_k \bar{d}_{\bar{k}} \prod_{k \in O} \bar{d}_k \ket{0_c}\\[5pt]
    & =\prod_{k \in P} \left(
        u_k + v_k \bar{c}_k^\dag \bar{c}_{\bar{k}}^\dag
    \right) \prod_{k \in O} \bar{c}_k^\dag \ket{0_c}.
\end{split}
\end{equation}
with $u_k,v_k\in \mathbb{R}$, $u_k^2+v_k^2=1$, and $|0\rangle_c$ the bare electron vacuum. 
The $\bar{c}_i$ and $\bar{d}_i$ operators are given by the transformations
\begin{equation}
    c_i = \sum_j C_{ij} \bar{c}_j,
    \quad
    d_i = \sum_j D_{ij} \bar{d}_j
    \label{eq:bar-basis}
\end{equation}
with $C(t)$ and $D(t)$ unitary transformations given such that the Boguliubov coherence matrices $U(t)$ and $V(t)$ are given by $U(t) = C(t) \bar{U}(t) D^\dag(t)$ and $V(t) = C^*(t) \bar{V}(t) D^\dag(t)$, with 

\begin{equation}
    \bar{U} = \begin{pmatrix}
        I & & \\
        & \oplus_k u_k \sigma_0 & \\
        & & 0
    \end{pmatrix},
    \quad
    \bar{V} = \begin{pmatrix}
        0 & & \\
        & \oplus_k v_k i\sigma_y & \\
        & & I
    \end{pmatrix}.
    \label{eq:BM}
\end{equation}
The overlap between two such time-dependent Fock-states $\ket{\vec{n}_d(t)}$, $\ket{\vec{n}'_d(t')}$ will then be given by

\begin{equation}
\begin{split}
   \braket{\vec{n}_d(t) |\vec{n}'_d(t')}
    &= \pm \frac{e^{i(\varphi(t') - \varphi(t))}}{\sqrt{\mathcal{N}(t) \mathcal{N}(t')}}
    \text{pf} \begin{pmatrix}
        \wick{\c {\bar{d}}^\dag(t) \c {\bar{d}}^\dag(t)} &
        \wick{\c {\bar{d}}^\dag(t) \c d(t)} &
        \wick{\c {\bar{d}}^\dag(t) \c d^\dag(t')} &
        \wick{\c {\bar{d}}^\dag(t) \c {\bar{d}}(t')} \\[3pt]
        &
        \wick{\c d(t) \c d(t)} &
        \wick{\c d(t) \c d^\dag(t')} &
        \wick{\c d(t) \c {\bar{d}}(t')} \\[3pt]
        &&
        \wick{\c d^\dag(t') \c d^\dag(t')} &
        \wick{\c d^\dag(t') \c {\bar{d}}(t')} \\[3pt]
        &&&
        \wick{\c {\bar{d}}(t') \c {\bar{d}}(t')}
    \end{pmatrix}\\[10pt] 
    &= \pm \frac{e^{i(\varphi(t') - \varphi(t))}}{\sqrt{\mathcal{N}(t) \mathcal{N}(t')}}
    \text{pf} \begin{pmatrix}
        \bar{V}^T(t)\bar{U}(t) &
        \bar{V}^T(t) C^\dag(t) V^*(t) &
        \bar{V}^T(t) C^\dag(t)  U(t') &
        \bar{V}^T(t) C^\dag(t) C(t') \bar{V}(t') \\[3pt]
        &
        U^\dag(t) V^*(t) &
        U^\dag(t) U(t') &
        U^\dag(t) C(t') \bar{V}(t')  \\[3pt]
        &&
        V^T(t') U(t') &
       V^T(t') C(t') \bar{U}(t') \\[3pt]
        &&&
        \bar{U}(t')  \bar{V}(t')
    \end{pmatrix}
\end{split}\end{equation}
with $\mathcal{N}(t) = \prod_{k \in P} v_k(t)^2$ and $\varphi(t)$, $\varphi(t')$ $\in [0,2\pi]$ phase factors. The method is used to calculate full many-body transition probabilities throughout this work, with Pfaffians calculated via the numerical routine given in \cite{wimmer2012}.





\bibliography{Hybrid_MZM}